\documentclass[onecolumn,draftcls]{IEEEtran}

\usepackage{amsfonts, amssymb, amsmath,,cite, enumerate,amsthm, bm, caption, subfig,psfrag,cases,mathtools,float}
\usepackage{ifthen}
\usepackage{flushend} 
\usepackage[usenames,dvipsnames]{color}
\ifCLASSINFOpdf
   \usepackage[pdftex]{graphicx}
   \DeclareGraphicsExtensions{.pdf,.jpeg,.png}
\else
   \usepackage[dvips]{graphicx}
   \DeclareGraphicsExtensions{.eps}
\fi

\DeclareMathOperator{\E}{\mathbb{E}}

\DeclareMathOperator*{\diag}{diag}

\DeclareMathOperator{\dist}{dist}

\DeclareMathOperator{\sign}{sign}

\newcommand{\ip}[2]{\left\langle#1,#2\right\rangle}
\newcommand{\norm}[1]{\left\lVert#1\right\rVert}
\renewcommand{\(}{\left(}
\renewcommand{\)}{\right)}
\renewcommand{\[}{\left[}
\renewcommand{\]}{\right]}

\newboolean{draft}
\newcommand{\isdraft}[2]{\ifthenelse{\boolean{draft}}{#1}{#2}}

\isdraft{\usepackage{setspace}}{}                              
\isdraft{\usepackage[footnotesize]{caption}}{}                 
\isdraft{\usepackage{paralist}}{}                              

\def \R {\mathbb{R}}

\def \S {\mathbb{S}}

\def \< {\langle}
\def \> {\rangle}

\def \vd {\bm{d}}
\def \ve {\bm{e}}
\def \vg {\bm{g}}
\def \vh {\bm{h}}
\def \vn {\bm{n}}
\def \vp {\bm{p}}
\def \vq {\bm{q}}

\def \vx {\bm{x}}
\def \vu {\bm{u}}

\def \vw {\bm{w}}
\def \vy {\bm{y}}
\def \vz {\bm{z}}
\def \vX {\bm{X}}
\def \vY {\bm{Y}}
\def \vA {\bm{A}}

\def \vG {\bm{G}}
\def \vW {\bm{W}}

\def \vV {\bm{V}}
\def \vU {\bm{U}}

\def \vZ {\bm{Z}}

\def \vphi {\bm{\phi}}

\def \vSigma {\bm{\Sigma}}

\def \vPhi   {\bm{\Phi}}

\theoremstyle{plain}
\newtheorem{theorem}{Theorem}

\newtheorem{fact}{Fact}

\newtheorem{lemma}{Lemma}

\theoremstyle{remark}
\newtheorem{remark}{Remark}

\begin{document}
\setboolean{draft}{true}
\title{Recovery of Structured Signals with Prior Information via Maximizing Correlation}

\author{Xu~Zhang, Wei~Cui, and Yulong~Liu
\thanks{The material in this paper was presented in part at the IEEE International Symposium on Information Theory, Aachen, Germany, 2017 \cite{zhang2017compressed}.}
\thanks{X.~Zhang and W.~Cui are with the School of Information and Electronics, Beijing Institute of Technology, Beijing 100081, China (e-mail:connorzx@bit.edu.cn; cuiwei@bit.edu.cn).}%
\thanks{Y.~Liu is with the School of Physics, Beijing Institute of Technology, Beijing 100081, China (e-mail: yulongliu@bit.edu.cn).}
}%

%



\maketitle

\begin{abstract}
This paper considers the problem of recovering a structured signal from a relatively small number of noisy measurements with the aid of a similar signal which is known beforehand. We propose a new approach to integrate prior information into the standard recovery procedure by maximizing the correlation between the prior knowledge and the desired signal. We then establish performance guarantees (in terms of the number of measurements) for the proposed method under sub-Gaussian measurements. Specific structured signals including sparse vectors, block-sparse vectors, and low-rank matrices are also analyzed. Furthermore, we present an interesting geometrical interpretation for the proposed procedure. Our results demonstrate that if prior information is good enough, then the proposed approach can (remarkably) outperform the standard recovery procedure. Simulations are provided to verify our results.

%
%
%
%
\end{abstract}


%
\IEEEpeerreviewmaketitle

\section{Introduction}

Compressed sensing (CS) concerns the problem of recovering a high-dimensional sparse (or nearly sparse) signal from a relatively small number of noisy measurements
\begin{equation}\label{Observation_model}
  \vy= \vA \vx^\star + \vn,
\end{equation}
where $\vA \in \R^{m \times n}$ is the measurement matrix, $\vx^\star \in \R^n$ denotes the signal to be estimated, and $\vn \in \R^m$ is the observation noise vector.  To reconstruct $\vx^\star$, a standard approach, named Lasso \cite{tibshirani1996regression} or Basis Pursuit (BP) \cite{chen2001atomic}, was proposed to use the $\ell_1$-norm as a surrogate function to promote sparsity constraint
\begin{equation}\label{eq: Classical_Problem}
    \min \limits_{\vx} \norm{\vx}_1 ~~~~\text{s.t.} ~~ \|\vy-\vA \vx\|_2 \leq \delta,
\end{equation}
where $\|\vx\|_p = \left(\sum_{i=1}^{n}|\vx_i|^p\right)^{1/p}$ denotes the standard $\ell_p$-norm of $\vx$, and $\delta$ is the upper bound (in terms of the $\ell_2$-norm) of the noise vector $\vn$. The performance of this method \eqref{eq: Classical_Problem} has been extensively studied in the literature, see e.g., \cite{donoho2006compressed,candes2006robust,Candes2006Near,eldar2012compressed,foucart2013mathematical,buhlmann2011statistics} and references therein.

However, in many practical applications, it is possible to have access to some prior knowledge about the desired signal in addition to the sparsity constraint. For instance, in video acquisition \cite{stankovi2009compressive,kang2009distributed,sankaranarayanan2015video}, dynamic system estimation \cite{ziniel2010tracking,charles2011sparsity}, and medical imaging \cite{lustig2007sparse,chen2008prior,weizman2015compressed}, past signals are very similar to the signal to be acquired, and hence might be utilized as prior information.
A natural question to ask is whether we can use these prior knowledge to improve the performance of the standard CS.

Within the past few years, there have been a variety of forms to integrate additional side information into the standard CS. For example, Vaswani and Lu \cite{vaswani2010modified}
 study the problem of recovering a sparse signal from a limited number of noiseless measurements when an estimate $T \subset \{1, \ldots, n\}$ of the support of $\vx^{\star}$ is known a priori. An exact reconstruction condition is established for the modified BP in which the objective in \eqref{eq: Classical_Problem} is replaced by $\norm{\vx_{T^c}}_1$, where $T^c$ is the complement of $T$ in $\{1, \ldots, n\}$ and $\vx_{T^c} \in \R^n$ denotes the vector that coincides with $\vx$ on $T^c$ and is set to zero on $T$. Their results have shown that if $T$ is a reasonable estimate of the support of $\vx^{\star}$, then the modified BP can improve the performance of the standard one.  \cite{vaswani2010modified} also considers an extended version of the modified BP but without any theoretical analysis
\begin{equation*}
  \min \limits_{\vx} \norm{\vx_{T^c}}_1 + \lambda \norm{\vx_{T}-\vphi_T}_2^2~~~~\text{s.t.}~~\vy=\vA\vx,
\end{equation*}
where $\vphi_{T}$ denotes an estimate of the components of $\vx^\star$ on $T$ and $\lambda$ is a tradeoff parameter.

Another line of work utilizes the known weights of the components of the desired signal as prior information. In \cite{tanaka2010optimal,khajehnejad2011analyzing,scarlett2013compressed}, a weighted $\ell_1$-norm minimization is proposed to recover the original signal
$$\min \limits_{\vx} \sum_{i=1}^{n} r_i\vx_i~~~~\text{s.t.}~~\vy=\vA\vx,$$
where $r_i>0$ are known weights. Their results have established that with suitable weights, the weighted BP has a better performance than the unweighted one.

In addition, a similar signal $\vphi$ of the original signal $\vx^\star$ can be naturally regarded as prior information.  The approaches in this group \cite{chen2008prior,weizman2015compressed,mota2017compressed} are to modify the objective function in BP by adding a new penalty term which penalizes the differences between $\vx^{\star}$ and the prior information $\vphi$, that is
\begin{equation}\label{eq: Original_Problem}
    \min \limits_{\vx} \norm{\vx}_1+ \lambda g(\vx-\vphi) ~~~~\text{s.t.} ~~  \|\vy-\vA \vx\|_2 \leq \delta,
\end{equation}
    where $\lambda > 0$ establishes a tradeoff between signal sparsity and fidelity to prior information, and $g: \R^n \rightarrow \R$ is a function which measures the similarity between $\vx^{\star}$ and $\vphi$. In \cite{mota2017compressed}, Mota et al. analyze two specific convex models for $g$: $g_1(\vx,\vphi)=\norm{\vx-\vphi}_1$ and $g_2(\vx,\vphi)=\norm{\vx-\vphi}_2^2/2$. Specifically, under Gaussian measurements, they derive the bounds on the number of measurements required to successfully recover the original signal for the following two models
\begin{equation}\label{eq: l1_l1_minimization}
    \min \limits_{\vx} \norm{\vx}_1+ \lambda \norm{\vx-\vphi}_1 ~~~~\text{s.t.} ~~  \|\vy-\vA \vx\|_2 \leq \delta,
\end{equation}
and
\begin{equation}\label{eq: l1_l2_minimization}
    \min \limits_{\vx} \norm{\vx}_1+  \frac{\lambda}{2} \norm{\vx-\vphi}_2^2~~~~\text{s.t.}~~\|\vy-\vA \vx\|_2 \leq \delta,
\end{equation}
which are called $\ell_1$-$\ell_1$ minimization and $\ell_1$-$\ell_2$ minimization, respectively. Their results have shown that, under suitable prior information, the former method improves the performance of CS largely while the latter brings no obvious benefits.

In this paper, we consider a new approach to incorporate prior information into the standard CS by maximizing the correlation between $\vx^{\star}$ and $\vphi$, which leads to
\begin{equation}\label{eq: Main_Problem L1norm}
    \min \limits_{\vx} \norm{\vx}_1 - \lambda \ip{\vx}{\vphi} ~~~~\text{s.t.} ~~  \|\vy-\vA \vx \|_2 \leq \delta,
\end{equation}
where $\lambda > 0$ is a tradeoff parameter. The motivation is natural because if $\vphi$ is similar to $\vx^{\star}$, then they may be highly correlated. Moreover, in addition to sparse signals, this framework can be easily applied to other structured signals. Therefore, we also generalize \eqref{eq: Main_Problem L1norm} to the following structured signal recovery procedure
\begin{equation}\label{eq: Main_Problem}
    \min \limits_{\vx} \norm{\vx}_{\text{sig}} - \lambda \ip{\vx}{\vphi} ~~~~\text{s.t.} ~~  \|\vy-\vA \vx \|_2 \leq \delta,
\end{equation}
where $\norm{\vx}_{\text{sig}}$ is a suitable norm which promotes the structure of the signal. Typical examples of structured signals include sparse vectors and low-rank matrices, which exhibit low complexity under the $\ell_1$-norm and the nuclear norm respectively. Additional examples of structured signals can be founded in \cite{chandrasekaran2012convex}.



We then establish the performance guarantees for the proposed method under sub-Gaussian measurements. More precisely, we establish the bound on the number of measurements required to successfully recover the original structured signal for the procedure \eqref{eq: Main_Problem}. We also apply the general result to specific signal structures of interest, including sparse vectors, block-sparse vectors, and low-rank matrices. In each case, we provide interpretable conditions under which the procedure \eqref{eq: Main_Problem} recovers the desired signal stably. Furthermore, we present an interesting geometrical interpretation for the proposed approach. Our results reveal that with proper prior information, our approach can (remarkably) outperform the standard recovery procedure.

%
%

The paper is organized as follows. In Section \ref{sec: Preliminaries}, we introduce some useful notations and facts which will be used in our analysis. Performance guarantees are presented in Section \ref{sec: Performance guarantees}. The geometrical interpretation for the proposed approach is given in Section \ref{sec: Geometrical_Interpretation}. Simulations are provided in Section \ref{sec: Simulation}, and conclusion is drawn in Section \ref{sec: Conclusion}.

\section{Preliminaries} \label{sec: Preliminaries}
In this section, we introduce some preliminaries which will be used in this paper. Throughout the paper, $\S^{n-1}$ denotes the unit sphere in $\R^n$ under the $\ell_2$-norm.

\subsection{Convex Geometry}
The \emph{subdifferential} of a convex function $f: \R^n \to \R$ at $\vx^\star$ is the set of vectors
\isdraft{$$
\partial f(\vx^\star) = \{\vu \in \R^n: f(\vx^\star + \vd) \geq f(\vx^\star) + \langle \vu, \vd \rangle~ \textrm{ for all}~\vd \in \R^n \}.
$$}{\begin{multline*}
  \partial f(\vx^\star) = \{\vu \in \R^n: f(\vx^\star + \vd) \geq f(\vx^\star) + \langle \vu, \vd \rangle~ \\ \textrm{ for all}~\vd \in \R^n \}.
\end{multline*}
}

The \emph{tangent cone} of a convex function $f$ at $\vx^\star$ is defined as
$$
    \mathcal{T}_f = \{\vd \in \R^n: f(\vx^\star+t \cdot \vd) \le f(\vx^\star)~~\textrm{for some}~t>0\},
$$
which is the set of descent directions of $f$ at $\vx^\star$.
The \emph{normal cone} of $f$ at $\vx^\star$ is the polar of the tangent cone
$$
\mathcal{N}_f= \{\vp \in \R^n: \ip{\vd}{\vp} \le 0 ~~ \textrm{for all}~\vd \in \mathcal{T}_f\}.
$$

The \emph{Gaussian width} and  \emph{Gaussian complexity} of a subset $\mathcal{E} \subset \R^n$ are defined as
$$
    w(\mathcal{E})= \E \sup \limits_{\vx \in \mathcal{E}} \ip{\vg}{\vx},~\vg \sim N(0,\bm{I}_n)
$$
and
$$
    \gamma (\mathcal{E})= \E \sup \limits_{\vx \in \mathcal{E}} |\ip{\vg}{\vx}|,~\vg \sim N(0,\bm{I}_n)
$$
respectively.
These two geometric quantities have a close relationship \cite{Chen2017}
\begin{equation}\label{Relation}
\gamma(\mathcal{E}) \le  2 w(\mathcal{E})+\norm{\vy}_2~~~\textrm{for every}~\vy \in \mathcal{E}.
\end{equation}

\subsection{Sub-Gaussian Random Vector}

A random variable $x$ is a \emph{sub-Gaussian random variable} if it has finite Orlicz norm
$$
\norm{x}_{\psi_2}=\inf \{t>0: \E \exp(x^2/t^2)\le 2\}.
$$
The sub-Gaussian norm of $x$, denoted $\norm{x}_{\psi_2}$, is the smallest $t$ such that $\E \exp(x^2/t^2)\le 2$. A random vector $\vx \in \R^n$ is called a \emph{sub-Gaussian random vector} if all of its one-dimensional marginals are sub-Gaussian, and its sub-Gaussian norm is defined as
$$
\norm{\vx}_{\psi_2} = \sup \limits_{\vy \in \S^{n-1}} \norm{\left\langle \vx,\vy \right\rangle}_{\psi_2}.
$$
We call a random vector $\vx \in \R^n$  \emph{isotropic} if it satisfies $\E \vx \vx^T = \bm{I}_n$, where $\bm{I}_n$ is the $n$-dimensional identity matrix.

\subsection{Some Useful Facts}

We also use the following well-established facts to derive our main results.

\begin{fact} [Theorem 1.3.5, VI, \cite{hiriart1993convex}] \label{Prop: NormalCone}
Let $f: \R^n \to \R$ be a convex function and suppose that $\bm{0}  \notin \partial f(\vx)$. Then the normal cone $\mathcal{N}_f$ of $f(\vx)$ at $\vx$ is
$$
    \mathcal{N}_f=\text{\emph{cone}}\{\partial f(\vx)\}.
$$
\end{fact}

\begin{fact}[Proposition 3.6,\cite{chandrasekaran2012convex}] \label{Prop: BoundGaussWidth}
  Let $\mathcal{T}$ be any nonempty convex cone in $\R^n$, and let $\vg \sim N(0,\bm{I}_n)$ be a random Gaussian vector. Then we have the following bound
  $$ w(\mathcal{T} \cap \S^{n-1}) \le \E[\dist(\vg,\mathcal{N})],
  $$
  where $\mathcal{N}$ is the polar of  $\mathcal{T}$ and $\dist(\vg,\mathcal{N}):= \min \{ \norm{\vg-\vh}_2:\vh \in \mathcal{N}\}. $
\end{fact}

\begin{fact}[Matrix deviation inequality, \cite{liaw2016simple}] \label{lm:MatrixDeviationInequality}
 Let $\vA$ be an ${m \times n}$ random matrix whose rows $\{\vA_{i\cdot}\}_{i=1}^m$ are independent, centered, isotropic and sub-Gaussian random vectors. For any bounded subset $\mathcal{D} \subset \R^n$ and $t \ge 0$, the event
$$
    \sup \limits_{\vx \in \mathcal{D}} \left| \norm{\vA \vx}_2 - \sqrt{m} \norm{\vx}_2 \right| \le CK^2 [\gamma (\mathcal{D})+t\cdot \text{\emph{rad}}(\mathcal{D})]
$$
holds with probability at least $1-2\exp(-t^2)$. Here $ \text{\emph{rad}}(\mathcal{D})=\sup_{\vx \in \mathcal{D}} \norm{\vx}_2$ and $K=\max_i \norm{\vA_{i\cdot}}_{\psi_2}$.
\end{fact}

\section{Performance guarantees} \label{sec: Performance guarantees}
In this section, we begin by establishing the performance guarantees for the proposed approach \eqref{eq: Main_Problem}. We then apply the general result to specific structured signals, including sparse vectors,
block-sparse vectors, and low-rank matrices. Our main results show that if prior information is good enough, the proposed approach can achieve a better performance than the standard recovery procedure.

%

\begin{theorem} \label{thm: Generalcase} Let $\vA \in \R^{m \times n}$ be a random matrix whose rows $\{\vA_{i\cdot}\}_{i=1}^m$ are independent, centered, isotropic and sub-Gaussian random vectors, and let $\mathcal{T}_f$ denote the tangent cone of $f(\vx)=\norm{\vx}_{\rm{sig}}-\lambda \ip{\vx}{\vphi}$. If
\begin{equation}\label{NumberofMeasurements}
        \sqrt{m} \ge CK^2 \gamma(\mathcal{T}_f \cap \S^{n-1})+ \epsilon,
\end{equation}
    then with probability at least $1- 2\exp(-\gamma^2(\mathcal{T}_f \cap \S^{n-1}))$, the solution $\hat{\vx}$ to \eqref{eq: Main_Problem} satisfies
    $$
    \norm{\hat{\vx}-\vx^\star}_2 \le \frac{2\delta}{\epsilon},
    $$
    where $\epsilon,C$ are absolute constants and $K=\max_i \norm{\vA_{i\cdot}}_{\psi_2}$.
\end{theorem}
\begin{IEEEproof}
Let $\vh = \hat{\vx} - \vx^\star$. Since $\hat{\vx}$ solves \eqref{eq: Main_Problem}, we have $\vh \in \mathcal{T}_f$ and $\vh/\norm{\vh}_2 \in \mathcal{T}_f \cap \S^{n-1}$. It follows from Fact \ref{lm:MatrixDeviationInequality} (let $\mathcal{D} = \mathcal{T}_f \cap \S^{n-1}$) that the event
\isdraft{
\begin{equation}\label{Restricted_Eigenvulue}
   \sqrt{m}- \inf \limits_{\bar{\vh} \in \mathcal{T}_f \cap \S^{n-1}} \norm{\vA \bar{\vh}}_2\leq \sup \limits_{\bar{\vh} \in \mathcal{T}_f \cap \S^{n-1}} \left| \norm{\vA \bar{\vh}}_2 - \sqrt{m} \right| \le C'K^2 [\gamma(\mathcal{T}_f \cap \S^{n-1})+t]
\end{equation}
}
{
\begin{equation}\label{Restricted_Eigenvulue}
 \begin{aligned}
   \sqrt{m}- \inf \limits_{\bar{\vh} \in \mathcal{T}_f \cap \S^{n-1}} \norm{\vA \bar{\vh}}_2 &\leq \sup \limits_{\bar{\vh} \in \mathcal{T}_f \cap \S^{n-1}} \left| \norm{\vA \bar{\vh}}_2 - \sqrt{m} \right| \\
    &\le C'K^2 [\gamma(\mathcal{T}_f \cap \S^{n-1})+t]
   \end{aligned}
\end{equation}
}
holds with probability at least $1- 2\exp(-t^2)$. Here we have used the facts that $\norm{\bar{\vh}}_2=1$ and $\text{rad}(\mathcal{T}_f\cap \S^{n-1})=1$. Choose $t = \gamma(\mathcal{T}_f \cap \S^{n-1})$ in \eqref{Restricted_Eigenvulue} and if $\sqrt{m} \ge CK^2 \gamma(\mathcal{T}_f \cap \S^{n-1})+ \epsilon$, we have that the event
\begin{align*}
 \inf \limits_{\bar{\vh} \in \mathcal{T}_f \cap \S^{n-1}}\norm{\vA \bar{\vh}}_2\geq  \sqrt{m}-2C'K^2\gamma (\mathcal{T}_f \cap \S^{n-1})\geq \epsilon
 \end{align*}
holds with probability at least $1- 2\exp(-\gamma^2(\mathcal{T}_f \cap \S^{n-1}))$, where $C = 2C'$. Therefore, with desired probability, we have
\begin{equation}\label{eq: 2.2}
\|\vA\vh\|_2 = \norm{\hat{\vx}-\vx^\star}_2 \norm{\vA \frac{\vh}{\norm{\vh}_2}}_2 \geq \epsilon \norm{\hat{\vx}-\vx^\star}_2.
\end{equation}

On the other hand, since both $\hat{\vx}$ and $\vx^\star$ are feasible, by the triangle inequality, we obtain
\begin{equation}\label{eq: 2.1}
\|\vA\vh\|_2 \leq \norm{\vy-\vA\vx^\star}_2+\norm{\vy-\vA\hat{\vx}}_2 \leq 2 \delta.
\end{equation}

Combining \eqref{eq: 2.2} and \eqref{eq: 2.1} completes the proof.

\end{IEEEproof}

\begin{remark} Actually, Theorem \ref{thm: Generalcase} holds for any convex function $f$. So it extends the result in \cite[Corollary 3.3]{chandrasekaran2012convex} from Gaussian measurements to sub-Gaussian measurements.
In particular, this result can also be used to analyze both $\ell_1$-$\ell_1$ minimization \eqref{eq: l1_l1_minimization} and $\ell_1$-$\ell_2$ minimization \eqref{eq: l1_l2_minimization} under sub-Gaussian measurements.
\end{remark}

\begin{remark} By the relationship between Gaussian width and Gaussian complexity \eqref{Relation}, the condition \eqref{NumberofMeasurements}
 can also be expressed in terms of Gaussian width $\sqrt{m} \ge CK^2 [2 \cdot w(\mathcal{T}_f \cap \S^{n-1})+1]+ \epsilon = C'K^2 w(\mathcal{T}_f \cap \S^{n-1})+ \epsilon $. The second inequality holds because in practical applications we usually have $w(\mathcal{T}_f \cap \S^{n-1})>0$.
\end{remark}

\begin{remark} In the noiseless setting where $\delta=0$, Theorem \ref{thm: Generalcase} entails exact recovery of $\vx^\star$ with probability at least $1- 2\exp(-\gamma^2(\mathcal{T}_f \cap \S^{n-1}))$ as long as $\sqrt{m} \geq  CK^2 \gamma(\mathcal{T}_f \cap \S^{n-1})$.
\end{remark}

\begin{remark} The bounded noise in Theorem \ref{thm: Generalcase} can be easily extended to the Gaussian case, since the Gaussian distribution is essentially bounded (namely, bounded with high probability).
\end{remark}

To make use of Theorem \ref{thm: Generalcase}, it is sufficient to bound $\gamma(\mathcal{T}_f \cap \S^{n-1})$ or $w(\mathcal{T}_f \cap \S^{n-1})$ for specific signal structure. Here, we consider three kinds of typical structured signals, namely, sparse vectors, block-sparse vectors, and low rank matrices. In each case, we provide interpretable conditions in terms of the number of measurements for stable recovery. The proofs of technical lemmas are included in Appendix \ref{Appdendix}.

\subsection{Sparse Vectors}

For sparse signal recovery, the signal norm in \eqref{eq: Main_Problem} becomes the $\ell_1$-norm:
\begin{equation} \label{eq: Main_Problem_Sparse}
    \min \limits_{\vx} \norm{\vx}_1-\lambda \ip{\vx}{\vphi} ~~~~\text{s.t.} ~~ \|\vy-\vA \vx\|_2 \leq \delta.
\end{equation}

Let $I=\{i:\vx_i^\star \neq 0\}$ be the support of $\vx^\star$ and $\mathcal{H}$ be the space of vectors in $\R^n$ with support only on $I$. Then the subdifferential of $\norm{\vx^\star}_1$ is given by
\begin{equation*}
  \partial \norm{\vx^\star}_1=\sign(\vx^\star)+\left\{\bm{\theta} \in \mathcal{H}^{\perp}: \norm{\bm{\theta}}_{\infty} \le 1\right\},
\end{equation*}
where $\sign(\cdot)$ denotes the sign function and $\mathcal{H}^{\perp}$ is the  orthogonal complement of $  \mathcal{H}$.

Define
\begin{equation}\label{parameter_v_1}
  v_1:=\max \limits_{\vw \in \partial{\norm{\vx^\star}_1}-\lambda \vphi} \norm{\vw}_2^2
\end{equation}
and
\begin{equation}\label{parameter_u_1}
u_1:=\norm{\sign(\vx^\star) -\lambda \vphi}_2^2.
\end{equation}
Then we have the following bound for $w^2(\mathcal{T}_f \cap \S^{n-1})$.

\begin{lemma} \label{lm: GaussianWidth_Sparse} Let $\vx^\star \in \R^n$ be an $s$-sparse vector and $\vphi \in \R^n$ be its prior information. Let $\mathcal{T}_f$ denote the tangent cone of $ f(\vx):=\norm{\vx}_1 - \lambda \ip{\vphi}{\vx}$ at $\vx^\star$. If $\bm{0} \notin \partial \norm{\vx^\star}_1 - \lambda \vphi$, then
\isdraft{
$$
    w^2(\mathcal{T}_f \cap \S^{n-1}) \le \min \left\{ n \cdot \left( 1 - \frac{n}{v_1} \cdot \frac{2}{\pi} \(1-\frac{s}{n}\)^2 \right), s+(n-s)u_1 \right\}.
$$
}{
\begin{multline*}
   w^2(\mathcal{T}_f \cap \S^{n-1})\\ \le \min \left\{ n \cdot \left( 1 - \frac{n}{v_1} \cdot \frac{2}{\pi} \(1-\frac{s}{n}\)^2 \right), s+(n-s)u_1 \right\}.
\end{multline*}
}
\end{lemma}

\begin{remark} If there is no prior information, i.e., $\lambda \vphi=\bm{0}$, our approach (\ref{eq: Main_Problem_Sparse}) reduces to BP. In this case, \eqref{parameter_v_1} reduces to $v_0:=\max \limits_{\vw \in \partial{\norm{\vx^\star}_1}} \norm{\vw}_2^2 = n$ and hence our first bound becomes
$$
 w^2(\mathcal{T}_0 \cap \S^{n-1}) \le n \cdot \left( 1 - \frac{2}{\pi} \(1-\frac{s}{n}\)^2 \right),
$$
which coincides with the result in \cite[equation (9)]{foygel2014corrupted}. Here, $\mathcal{T}_0$ denotes the tangent cone of $\norm{\vx}_1$ at $\vx^\star$. However, if we choose some suitable $\lambda \vphi$ such that $v_1 \le v_0=n$, then the approach (\ref{eq: Main_Problem_Sparse}) can achieve a better performance than BP.
\end{remark}

\begin{remark} \label{rm: PreciseBound_Sparse_nonclosed}
The proof of Lemma \ref{lm: GaussianWidth_Sparse} implies a sharp bound for $w^2(\mathcal{T}_f \cap \S^{n -1})$. Indeed, it follows from \eqref{optimalboundequation} that
\begin{align}\label{OptimalBound}
  w^2(\mathcal{T}_f \cap \S^{n -1}) & \leq \E \left[\inf_{\vw \in \mathcal{N}_f} \|\vg - \vw\|_2^2\right] \\ \notag
                                    & =    \E \left[\min_{t\geq 0} \inf_{\vw \in  t \cdot (\partial \norm{\vx^\star}_1 - \lambda \vphi)}  \|\vg - \vw\|_2^2 \right]   \\ \notag
                                    & \leq \min_{t\geq 0} \E \left[\inf_{\vw \in  t \cdot (\partial \norm{\vx^\star}_1 - \lambda \vphi)}  \|\vg - \vw\|_2^2 \right]   \\ \notag
                                    & = \min_{t \ge 0}  \E \[ \dist^2(\vg, t \cdot (\partial \norm{\vx^\star}_1 - \lambda \vphi))\].
\end{align}
Clearly, this bound is tighter than the two ones (obtained with two fixed values of $t$) in Lemma \ref{lm: GaussianWidth_Sparse}, albeit without an interpretable closed form.
\end{remark}

\begin{remark} With proper prior information, the first bound closely approximates the optimal bound \eqref{OptimalBound} for a wide range of sparsity levels, while the second bound dominates the first one in the extreme sparsity regime. To see this, it follows from \eqref{parameter_v_1} that
\begin{equation*}
    v_1= \sum \limits_{i \in I} (\sign(\vx_i^\star)-\lambda \vphi_i)^2 + \sum \limits_{i \in I^c} (1+|\lambda \vphi_i|)^2 \ge  n-s,
\end{equation*}
and hence
$$
    n\left(1 - \frac{n}{v_1} \cdot \frac{2}{\pi} \(1-\frac{s}{n}\)^2 \right) \ge n- \frac{2}{\pi}(n-s).
$$
In the extreme sparsity regime, we have $n- 2(n-s)/{\pi} \gg s$, which explains why the first bound performs badly in this regime. However, for the second bound, suitable prior information can lead to $u_1 \to 0$ and $(n-s)u_1+s \to s$.
\end{remark}

Combining Theorem \ref{thm: Generalcase} with Lemma \ref{lm: GaussianWidth_Sparse}, we arrive at the following theorem.

\begin{theorem} \label{thm: SparseCase}Let $\vA$ be an ${m \times n}$ matrix whose rows are independent, centered, isotropic and sub-Gaussian random vectors and $\vx^\star \in \R^n$ be an $s$-sparse vector. Suppose that $ \bm{0} \notin  \partial \norm{\vx^\star}_1 -\lambda \vphi$. If
\isdraft{
    $$
        \sqrt{m} \ge CK^2 \min \left\{ \sqrt{n \cdot \left( 1 - \frac{n}{v_1} \cdot \frac{2}{\pi} \(1-\frac{s}{n}\)^2 \right)}, \sqrt{s+(n-s)u_1}\right\}+ \epsilon,
    $$
    }{
    \begin{multline*}
       \sqrt{m} \ge CK^2 \min \bigg\{ \sqrt{n \cdot \left( 1 - \frac{n}{v_1} \cdot \frac{2}{\pi} \(1-\frac{s}{n}\)^2 \right)}, \\
       \sqrt{s+(n-s)u_1} \bigg\} + \epsilon,
    \end{multline*}
    }
    then with probability $1- o(1)$, the solution $\hat{\vx}$ to (\ref{eq: Main_Problem_Sparse}) satisfies
    $$
    \norm{\hat{\vx}-\vx^\star}_2 \le \frac{2\delta}{\epsilon},
    $$
    where $\epsilon,C$ are absolute constants and $K=\max_i \norm{\vA_{i\cdot}}_{\psi_2}$.
\end{theorem}

\subsection{Block Sparse Vectors}
Let $\vx^{\star} \in \R^n$ be a block-sparse vector. Partition the indices $\{1,2,\ldots,n\}$ into $l$ disjoint blocks $V_1,\ldots,V_l$ of size $k$ and suppose that $\vx^{\star}$ is supported on at most $s$ blocks of them. After incorporating prior information $\vphi \in \R^n$ into the standard block-sparse recovery procedure, we get the following approach:
\begin{equation}\label{eq: Main_Problem_BlockSparse}
    \min \limits_{\vx} \norm{\vx}_{2,1} - \lambda \ip{\vx}{\vphi}~~~~\text{s.t.} ~~  \|\vy-\vA \vx\|_2 \leq \delta,
\end{equation}
where $\norm{\vx}_{2,1}=\sum_{i=1}^{l}\norm{\vx_{V_i}}_2$ and $\lambda>0$ is a tradeoff parameter.
 Let $\mathcal{B}$ be the $s$ block indices on which $\vx^\star$ is supported and $\mathcal{K}$ be the space of vectors in $\R^n$ with support only on $\mathcal{B}$. We have the subdifferential \cite{friedman2010note}
 \begin{equation*}
   \partial \norm{\vx^\star}_{2,1}= \sum_{b \in \mathcal{B}} \frac{\vx^\star_{V_b}}{\norm{\vx^\star_{V_b}}_2}+\left\{\bm{\theta} \in \mathcal{K}^\perp : \max \limits_{b \notin \mathcal{B}} \norm{\bm{\theta}_{V_b}}_2 \le 1 \right\},
 \end{equation*}
where $\vx^\star_{V_b} \in \R^n$ denotes the vector that coincides with $\vx$ on the block $V_b$ and is set to zero on the other blocks.

Define
\begin{equation}\label{parameter_v_2}
v_2:= \max \limits_{\vw \in \partial \norm{\vx^\star}_{2,1} - \lambda \vphi} \norm{\vw}_2^2
\end{equation}
and
\begin{equation}\label{parameter_u_2}
u_2: = \norm{\sum_{b \in \mathcal{B}} \frac{\vx_{V_b}^\star}{\norm{\vx_{V_b}^\star}_2}-\lambda \vphi}_2^2.
\end{equation}
Then we can bound $w^2(\mathcal{T}_f \cap \S^{n-1})$ as follows.

\begin{lemma} \label{lm: GaussianWidth_Blocksparse} Let $\vx^\star \in \R^n$ be an $s$-block-sparse vector and $\vphi \in \R^n$ be its prior information. Suppose that the indices $\{1,2,\ldots,n\}$ can be partioned into disjoint blocks $V_1,\ldots,V_l$ of size $k=n/l$. Let $\mathcal{T}_f$ denote the tangent cone of $ f(\vx):=\norm{\vx}_{2,1} - \lambda \ip{\vphi}{\vx}$ at $\vx^\star$. Suppose that $\bm{0} \notin \partial \norm{\vx^\star}_{2,1} - \lambda \vphi$, then
\isdraft{
$$
    w^2(\mathcal{T}_f \cap \S^{n-1}) \le \min\left\{n \cdot \left( 1 - \frac{l}{v_2} \cdot \frac{\mu_k^2}{k} \(1-\frac{s}{l}\)^2 \right),k\cdot \(s+(l-s)u_2\)\right\},
$$
}{\begin{multline*}
    w^2(\mathcal{T}_f \cap \S^{n-1}) \le \min\bigg\{n \cdot \left( 1 - \frac{l}{v_2} \cdot \frac{\mu_k^2}{k} \(1-\frac{s}{l}\)^2 \right), \\ k\cdot \(s+(l-s)u_2\)\bigg\},
  \end{multline*}
}
where $\mu_k$ is the mean of $\chi_k$ distribution.
\end{lemma}

\begin{remark}
If there is no prior information, i.e., $\lambda \vphi= \bm{0}$, we have $v_2=l$ and the first bound reduces to
\begin{equation*}
    w^2(\mathcal{T}_f \cap \S^{n-1}) \le
    n \cdot \left( 1 -\frac{\mu_k^2}{k} \(1-\frac{s}{l}\)^2 \right),
\end{equation*}
which coincides with the result in  \cite[Proposition 4 of Appendix A]{foygel2014corrupted}.
\end{remark}

Combining Theorem \ref{thm: Generalcase} with Lemma \ref{lm: GaussianWidth_Blocksparse}, we arrive at the following theorem.

\begin{theorem} \label{thm: BlockSparseCase} Let $\vA$ be an ${m \times n}$ matrix whose rows are independent, centered, isotropic and sub-Gaussian random vectors. With the assumptions of Lemma \ref{lm: GaussianWidth_Blocksparse}, if
\isdraft{         
    $$
        \sqrt{m} \ge CK^2 \min\left\{\sqrt{n \cdot \left( 1 - \frac{l}{v_2} \cdot \frac{\mu_k^2}{k} \(1-\frac{s}{l}\)^2 \right)},\sqrt{k\cdot \(s+(l-s)u_2\)}\right\}+ \epsilon,
    $$
}{                
    \begin{multline*}
      \sqrt{m} \ge CK^2 \min \bigg\{\sqrt{n \cdot \left( 1 - \frac{l}{v_2} \cdot \frac{\mu_k^2}{k} \(1-\frac{s}{l}\)^2 \right)}, \\
      \sqrt{k\cdot \(s+(l-s)u_2\)}\bigg\}+ \epsilon,
    \end{multline*}
}
    then with probability $1- o(1)$, the solution $\hat{\vx}$ to (\ref{eq: Main_Problem_BlockSparse}) satisfies
    $$
    \norm{\hat{\vx}-\vx^\star}_2 \le \frac{2\delta}{\epsilon},
    $$
    where $\epsilon,C$ are absolute constants and $K=\max_i \norm{\vA_{i\cdot}}_{\psi_2}$.
\end{theorem}

\subsection{Low Rank Matrices}
Let $\vX^\star\in \R^{n_1 \times n_2}$ be a rank $r$ matrix and $\vA \in \R^{n_1 \times n_2}$ be a measurement matrix whose rows are independent, isotropic, centered and sub-Gaussian vectors.  Without loss of generality, we suppose  $n_1 \ge n_2$. We want to recover $\vX^\star$ from the measurements $\vy =\mathcal{A} ( \vX^\star) + \vn \in \R^{m}$, where $\mathcal{A}: \vX \to \sum_{j=1}^{m} \ip{\vA^j}{\vX} \ve_j$. Here $\ve_1,\ldots,\ve_m$ denote the standard basis vectors in $\R^{m}$ and $\vA^1,\ldots,\vA^m$ are independent copies of $\vA$. We incorporate prior information $\vPhi$ into the standard low rank matrix recovery procedure, which leads to
\begin{equation}\label{eq: Main_Problem_LowRank}
\min \limits_{\vX} \norm{\vX}_*-\lambda \ip{\vPhi}{\vX} ~~~~\text{s.t.} ~~ \norm{\vy - \mathcal{A}(\vX)}_2 \le \delta,
\end{equation}
where $\norm{\vX}_*$ denotes the nuclear norm of $\vX$, which is equal to the sum of singular values of $\vX$, and $\lambda $ is a tradeoff parameter.

Let $\vX^\star=\vU \bm{\Sigma} \vV^T$ be the compact singular value decomposition (SVD) of $\vX^\star$. Let $\tilde{\vU}=[\vU,\vU']$ and $\tilde{\vV}=[\vV,\vV']$ be orthonormal matrices with $\vU \in \R^{n_1 \times r},\vU' \in \R^{n_1 \times (n_1-r)},\vV \in \R^{n_2 \times r}, $ and $\vV' \in \R^{n_2 \times (n_2-r)}$. Let $\vU_{\cdot k}$  and $\vV_{\cdot k}$ be the $k$-th column of $\vU$ and $\vV$, respectively. It is convenient to introduce the orthogonal decomposition: $\R^{n_1\times n_2}=\mathcal{S}+\mathcal{S}^{\bot}$, where $\mathcal{S}$ is the space spanned by elements of the form $\vU_{\cdot k} \vp^T$ and $\vq \vV_{\cdot k}^T$, $1\le k \le r$, where $\vp \in \R^{n_2}$ and $\vq \in \R^{n_1}$ are arbitrary vectors, and $\mathcal{S}^{\bot}$ is the orthogonal complement of $\mathcal{S}$ in $\R^{n_1\times n_2}$ \cite{candes2012exact}. Then for any matrix $\vX \in \R^{n_1 \times n_2}$, the orthogonal projection $\mathcal{P}_{\mathcal{S}}$ onto $\mathcal{S}$ is
$$\mathcal{P}_{\mathcal{S}}(\vX)=\vU\vU^T \vX + \vX \vV \vV^T- \vU \vU^T \vX \vV \vV^T,$$
and the orthogonal projection $\mathcal{P}_{\mathcal{S}^{\bot}}$ onto $\mathcal{S}^{\bot}$ is
$$\mathcal{P}_{\mathcal{S}^{\bot}}(\vX)=\vU'\vU'^T \vX \vV' \vV'^T.$$

We have known from \cite{Watson1992Characterization} that the subdifferential of  $\norm{\vX^\star}_*$ is given by
$$
\partial \norm{\vX^\star}_*=\vU \vV^T+\left\{\vW:\vW^T\vU=\bm{0}, \vW\vV=\bm{0}, \norm{\vW} \le 1 \right\},
$$
where $\norm{\vW}$ is the spectral norm of $\vW$.

Define
\begin{equation}\label{parameter_v_3}
   v_3:=\max \limits_{\vZ \in \partial \norm{\vX^\star}_* -\lambda \vPhi} \norm{\vZ}_F^2
\end{equation}
 and
\begin{equation}\label{parameter_u_3}
   u_3:=\norm{\vU\vV^T-\lambda \vPhi}_F^2.
\end{equation}
Then we have the following bound for $w^2(\mathcal{T}_f \cap \S^{n_1 n_2 -1})$.

\begin{lemma} \label{lm: GaussianWidth_LowRank} Let $\vX^\star \in \R^{n_1 \times n_2}$ be a rank $r$ matrix and $\vPhi \in \R^{n_1 \times n_2}$ be its prior information. Let $\mathcal{T}_f$ denote the tangent cone of $ f(\vX):=\norm{\vX}_*-\lambda \ip{\vPhi}{\vX}$ at $\vX^\star$. Suppose that $\bm{0} \notin \partial \norm{\vX^\star}_* - \lambda \vPhi$, then
\isdraft{               
\begin{multline*}
  w^2(\mathcal{T}_f \cap \S^{n_1 n_2 -1})
  \le \min \left\{n_1 n_2\( 1- \frac{n_2}{v_3} \(\frac{4}{27}\)^2 \(1-\frac{r}{n_1}\)\(1-\frac{r}{n_2}\)^2 \), \right.\\\left.
  r(n_1+n_2-r) + u_3\((\sqrt{n_1-r}+\sqrt{n_2-r})^2+2\)\right\}.
\end{multline*}
}{                      
\begin{multline*}
  w^2(\mathcal{T}_f \cap \S^{n_1 n_2 -1}) \\
  \le \min \Bigg\{n_1 n_2\( 1- \frac{n_2}{v_3} \(\frac{4}{27}\)^2 \(1-\frac{r}{n_1}\)\(1-\frac{r}{n_2}\)^2 \), \\ r(n_1+n_2-r) + u_3 \((\sqrt{n_1-r}+\sqrt{n_2-r})^2+2\) \Bigg\}.
\end{multline*}}
\end{lemma}

\begin{remark}
If there is no prior information, i.e., $\lambda \vPhi= \bm{0}$, we have $v_3=n_2$ and the first bound reduces to
\isdraft{               
\begin{equation*}
   w^2(\mathcal{T}_f \cap \S^{n_1 n_2 -1}) \le n_1n_2\( 1-\(\frac{4}{27}\)^2 \(1-\frac{r}{n_1}\)\(1-\frac{r}{n_2}\)^2 \),
\end{equation*}
}{                      
\begin{multline*}
    w^2(\mathcal{T}_f \cap \S^{n_1 n_2 -1}) \\
    \le n_1n_2\( 1-\(\frac{4}{27}\)^2 \(1-\frac{r}{n_1}\)\(1-\frac{r}{n_2}\)^2 \),
\end{multline*}}
which coincides with the result in  \cite[Proposition 5]{foygel2014corrupted}.
\end{remark}

\begin{remark} \label{rm: PreciseBound_LowRank_nonclosed}
Similar to Remark \ref{rm: PreciseBound_Sparse_nonclosed}, we can also obtain a sharp bound for $w^2(\mathcal{T}_f \cap \S^{n_1 n_2 -1})$
\begin{equation*}\label{OptimalBoundLowrank}
  w^2(\mathcal{T}_f \cap \S^{n_1 n_2 -1}) \leq \min_{t \ge 0} \E[\dist^2 (\vG,t(\partial \norm{\vX^\star}_* -\lambda \vPhi))].
\end{equation*}
This bound is tighter than the two ones in Lemma \ref{lm: GaussianWidth_LowRank}, but lack of an interpretable closed form.

\end{remark}

Combining Theorem \ref{thm: Generalcase} with Lemma \ref{lm: GaussianWidth_LowRank}, we arrive at the following theorem.
\begin{theorem} \label{thm: LowRankCase}Let $\vA$ be an ${n_1 \times n_2}$ matrix whose rows are independent, centered, isotropic and sub-Gaussian random vectors and $\vA^1,\ldots,\vA^m$ be its independent copies.
Let $\vX^\star \in \R^{n_1 \times n_2}$ be a rank $r$ matrix. Assume that $\mathcal{A}: \vX \to \sum_{j=1}^{m} \ip{\vA^j}{\vX} \ve_j$ and $ \bm{0} \notin  \partial \norm{\vX^\star}_* -\lambda \vPhi$. If
\isdraft{
\begin{multline*}
        \sqrt{m} \ge CK^2 \cdot \min \Bigg\{\sqrt{n_1 n_2\( 1- \frac{n_2}{v_3} \(\frac{4}{27}\)^2 \(1-\frac{r}{n_1}\)\(1-\frac{r}{n_2}\)^2 \)}, \\
  \sqrt{r(n_1+n_2-r) + u_3 \((\sqrt{n_1-r}+\sqrt{n_2-r})^2+2\)} \Bigg\}+ \epsilon,
    \end{multline*}
}{
    \begin{multline*}
        \sqrt{m} \ge CK^2 \cdot \\ \min \Bigg\{\sqrt{n_1 n_2\( 1- \frac{n_2}{v_3} \(\frac{4}{27}\)^2 \(1-\frac{r}{n_1}\)\(1-\frac{r}{n_2}\)^2 \)}, \\
  \sqrt{r(n_1+n_2-r) + u_3 \((\sqrt{n_1-r}+\sqrt{n_2-r})^2+2\)} \Bigg\}+ \epsilon,
    \end{multline*}
}
    then with probability $1- o(1)$, the solution $\hat{\vX}$ to \eqref{eq: Main_Problem_LowRank} satisfies
    $$
    \norm{\hat{\vX}-\vX^\star}_F \le \frac{2\delta}{\epsilon},
    $$
    where $K = \norm{\text{vec}[(\vA^T)]}_{\psi_2}$ \footnote{Here $\text{vec}[\vA]$ denotes the column vector obtained by stacking the columns of the matrix $\vA$ on top of one another.} and $\epsilon,C$ are absolute constants.
\end{theorem}

\section{Geometrical Interpretation} \label{sec: Geometrical_Interpretation}

In this section, we present an interesting geometrical interpretation for our procedure \eqref{eq: Main_Problem} and suggest some strategies to improve the quality of prior information.

 Our main result (Theorem \ref{thm: Generalcase}) has shown that the number of measurements required for successful reconstruction is determined by the spherical Gaussian width of the tangent cone of $f(\vx)$ at $\vx^\star$, i.e., $w(\mathcal{T}_f \cap \S^{n-1})$. Recall that the normal cone $\mathcal{N}_f$ of $f(\vx)$ at $\vx^\star$ is the polar of its tangent cone $\mathcal{T}_f$. This implies that the larger the normal cone $\mathcal{N}_f$, the less the number of measurements required for successful recovery.

In the standard structured signal recovery procedure \cite{chandrasekaran2012convex}
\begin{equation}\label{eq: Main_Problem standard}
    \min \limits_{\vx} \norm{\vx}_{\text{sig}} ~~~~\text{s.t.} ~~  \|\vy-\vA \vx \|_2 \leq \delta,
\end{equation}
the subdifferential of the objective $\norm{\vx}_{\text{sig}}$ at $\vx^\star$ is $\partial \norm{\vx^\star}_{\text{sig}}$. If $\bm{0} \notin \partial \norm{\vx^\star}_{\text{sig}}$, then the corresponding normal cone is $\mathcal{N}_0 = \text{cone}\{\partial \norm{\vx^\star}_{\text{sig}}\}$. For the proposed recovery procedure (\ref{eq: Main_Problem}), the subdifferential of the objective $f(\vx)= \norm{\vx}_{\text{sig}} - \lambda \langle \vx, \vphi\rangle$ at $\vx^\star$ is $\partial \norm{\vx^\star}_{\text{sig}}- \lambda \vphi$, which is a shifted version of $\partial \norm{\vx^\star}_{\text{sig}} $. If $\bm{0} \notin \partial \norm{\vx^\star}_{\text{sig}} - \lambda \vphi$, then $\mathcal{N}_f= \text{cone}\{\partial \norm{\vx^\star}_{\text{sig}} - \lambda \vphi\}$. In order to achieve a better performance than the standard procedure, it is required that the prior information $\lambda \vphi$ is good enough such that $\mathcal{N}_f$ is larger than $\mathcal{N}_0$. This will lead to a smaller tangent cone and hence less number of measurements required for successful recovery.
\begin{figure}[H]
	\centering
	\subfloat[$\lambda_a\vphi_a=\[0,0\]^T$]{\includegraphics[width=3.0in]{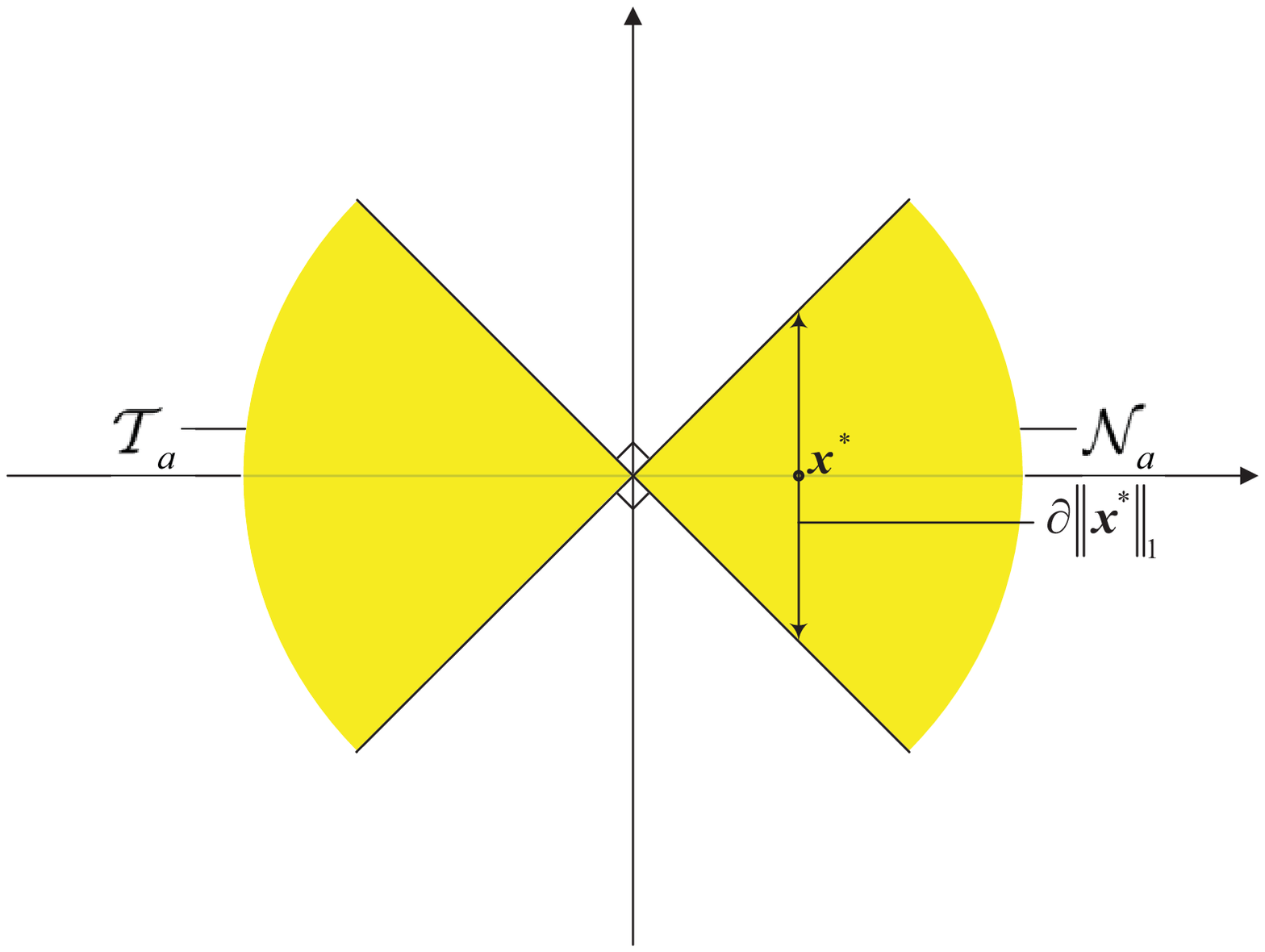}
		\label{fig_first_case}}
	\hfil
	\subfloat[$\lambda_b\vphi_b=\[0.5,0\]^T$]{\includegraphics[width=3.0in]{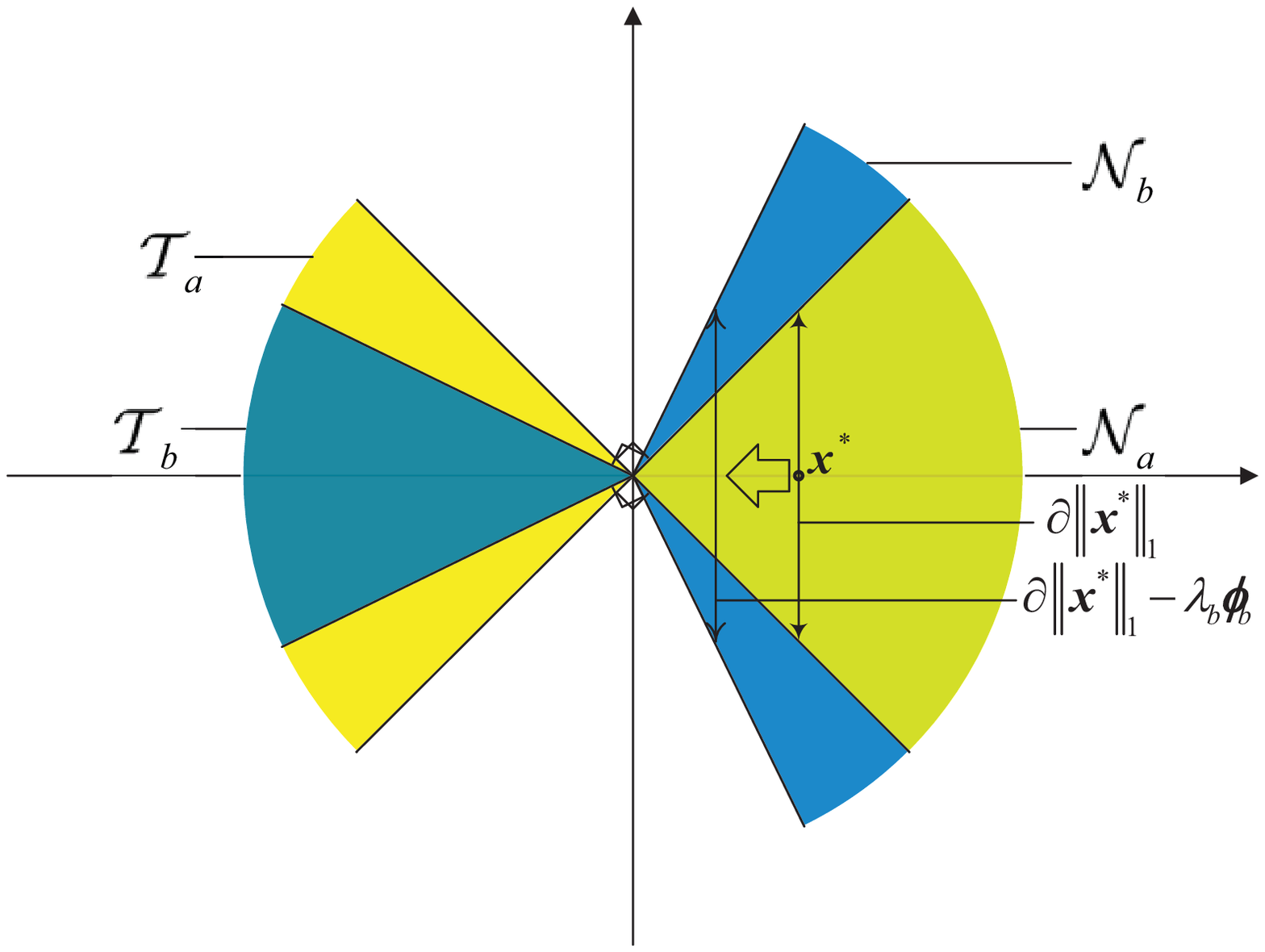}
		\label{fig_second_case}}
	\hfil
	\subfloat[$\lambda_c\vphi_c=\[-0.5,0\]^T$]{\includegraphics[width=3.0in]{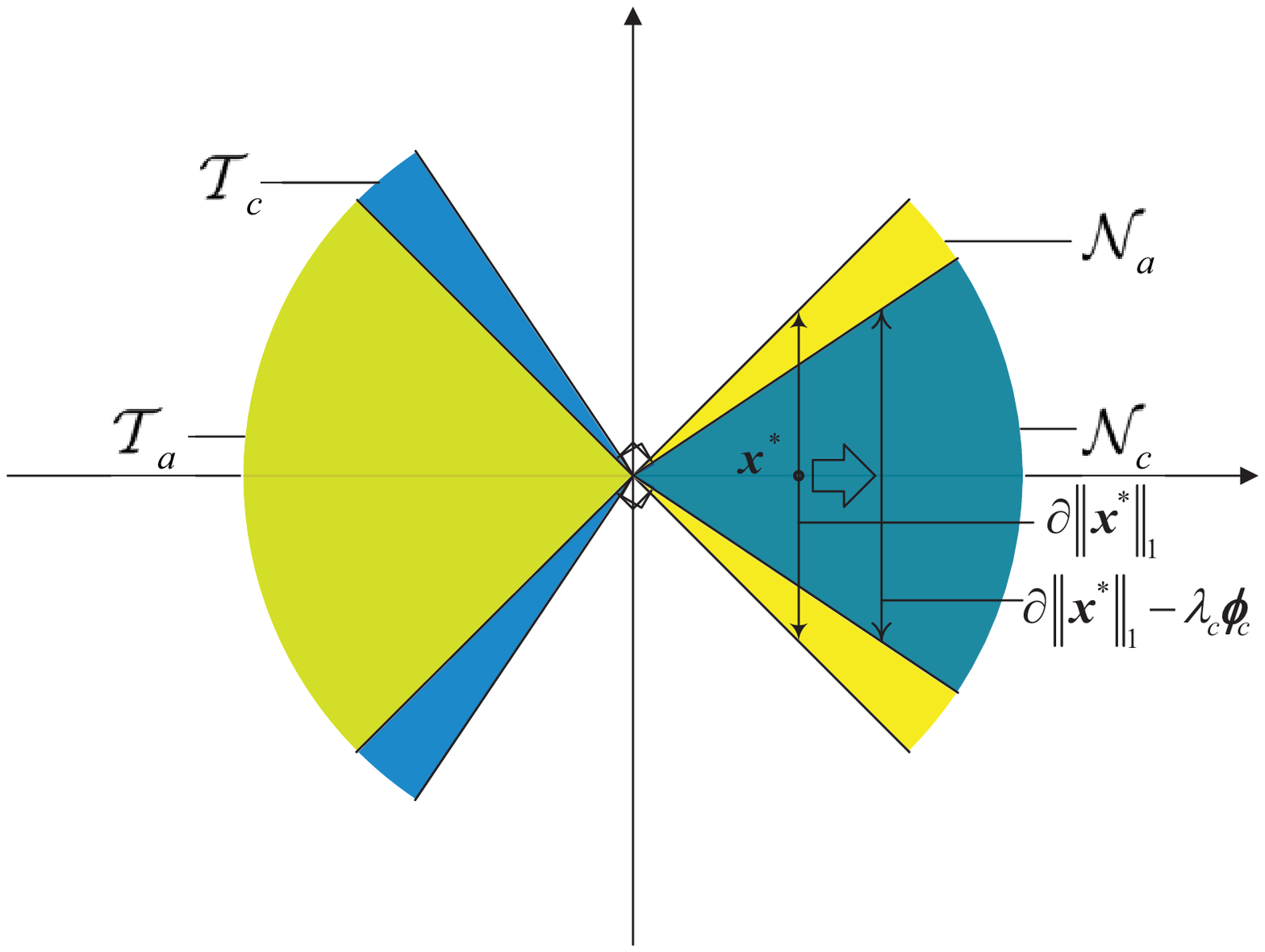}
		\label{fig_third_case}}
	\hfil
	\subfloat[$\lambda_d\vphi_d=\[0,-1\]^T$]{\includegraphics[width=3.0in]{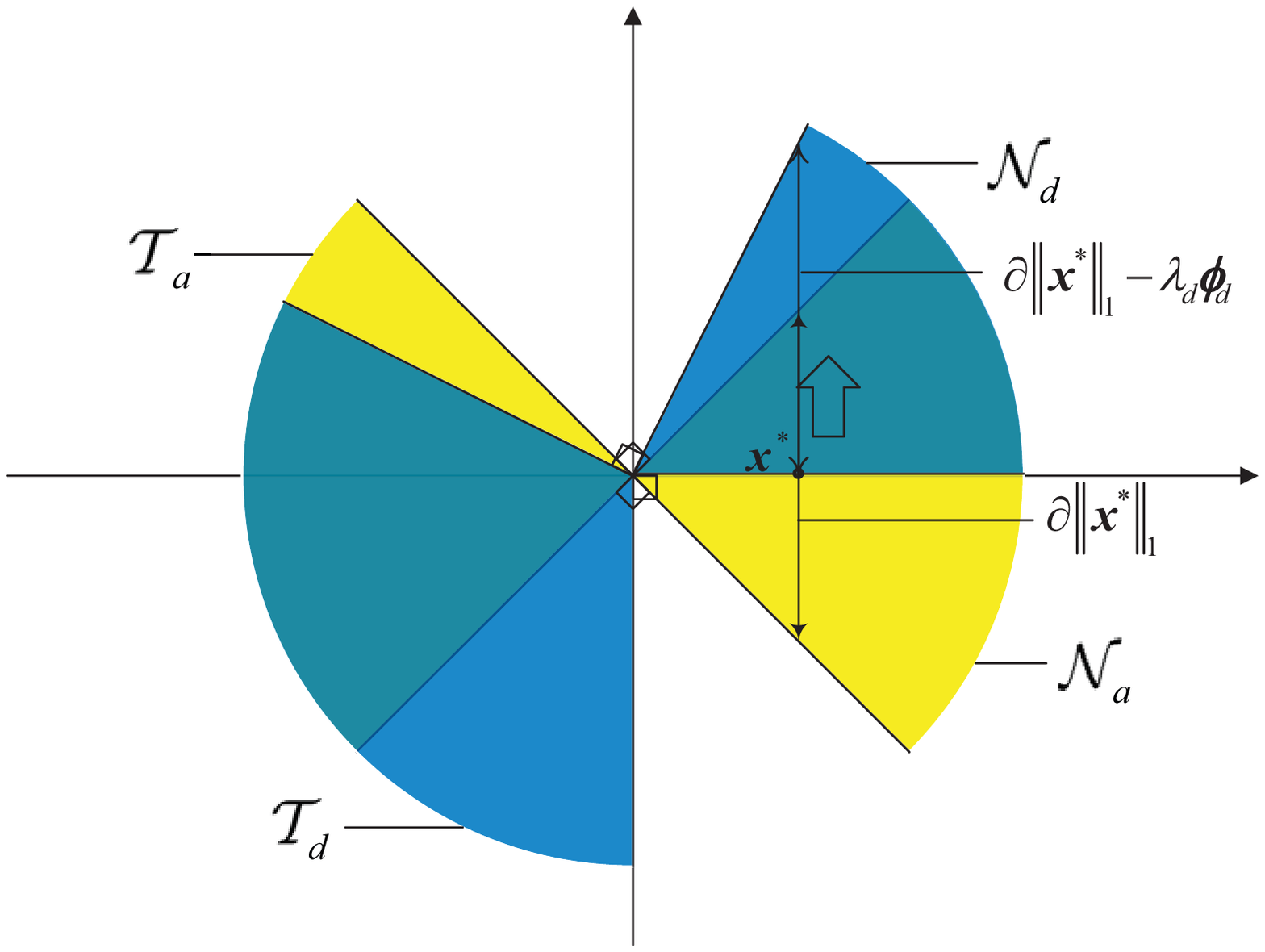}
		\label{fig_fourth_case}}
	\hfil
	\subfloat[$\lambda_e\vphi_e=\[0.5,-0.2\]^T$]{\includegraphics[width=3.0in]{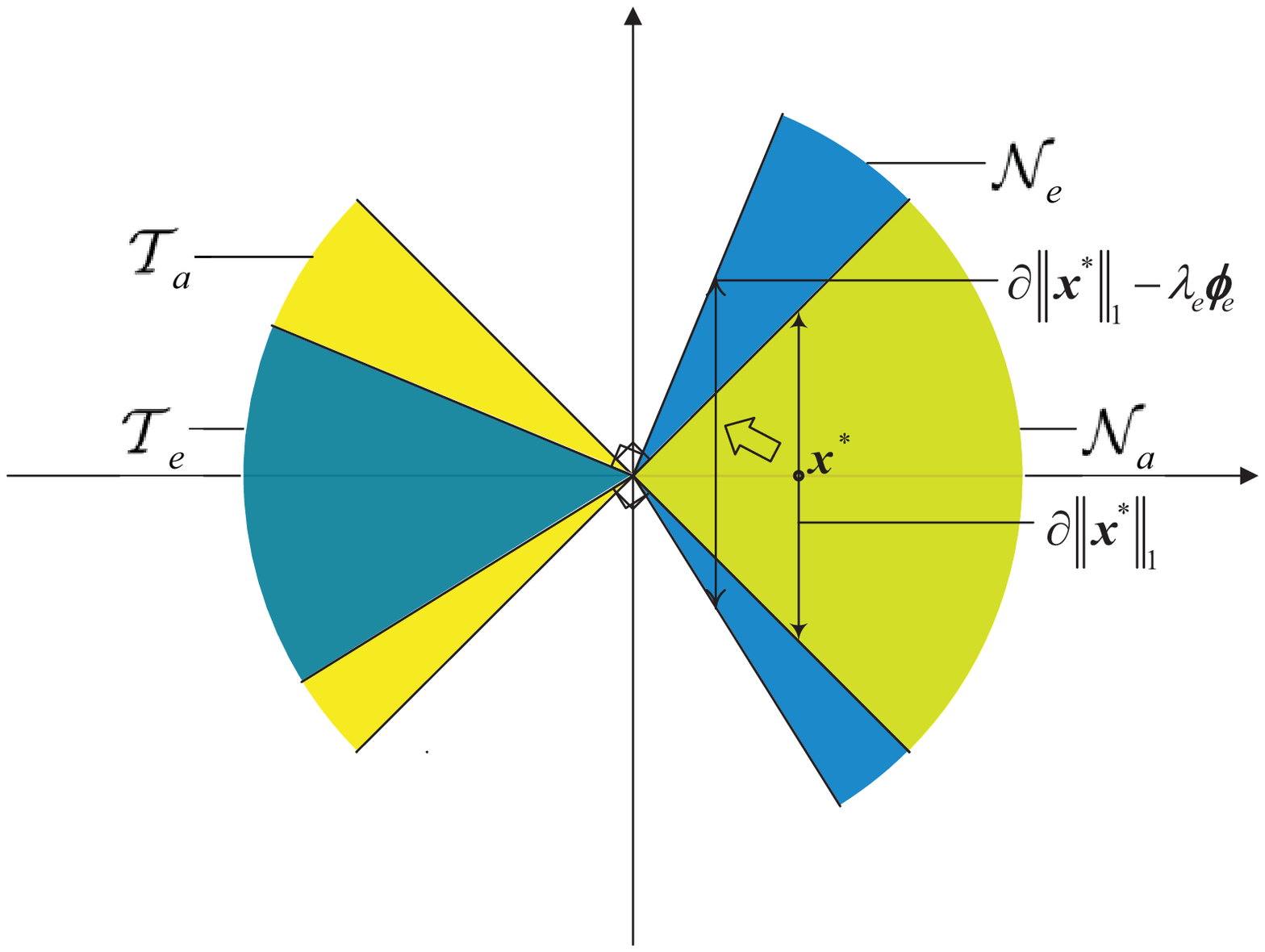}
		\label{fig_fifth_case}}
	\hfil
	\subfloat[$\lambda_f\vphi_f=\[0.5,-1\]^T$]{\includegraphics[width=3.0in]{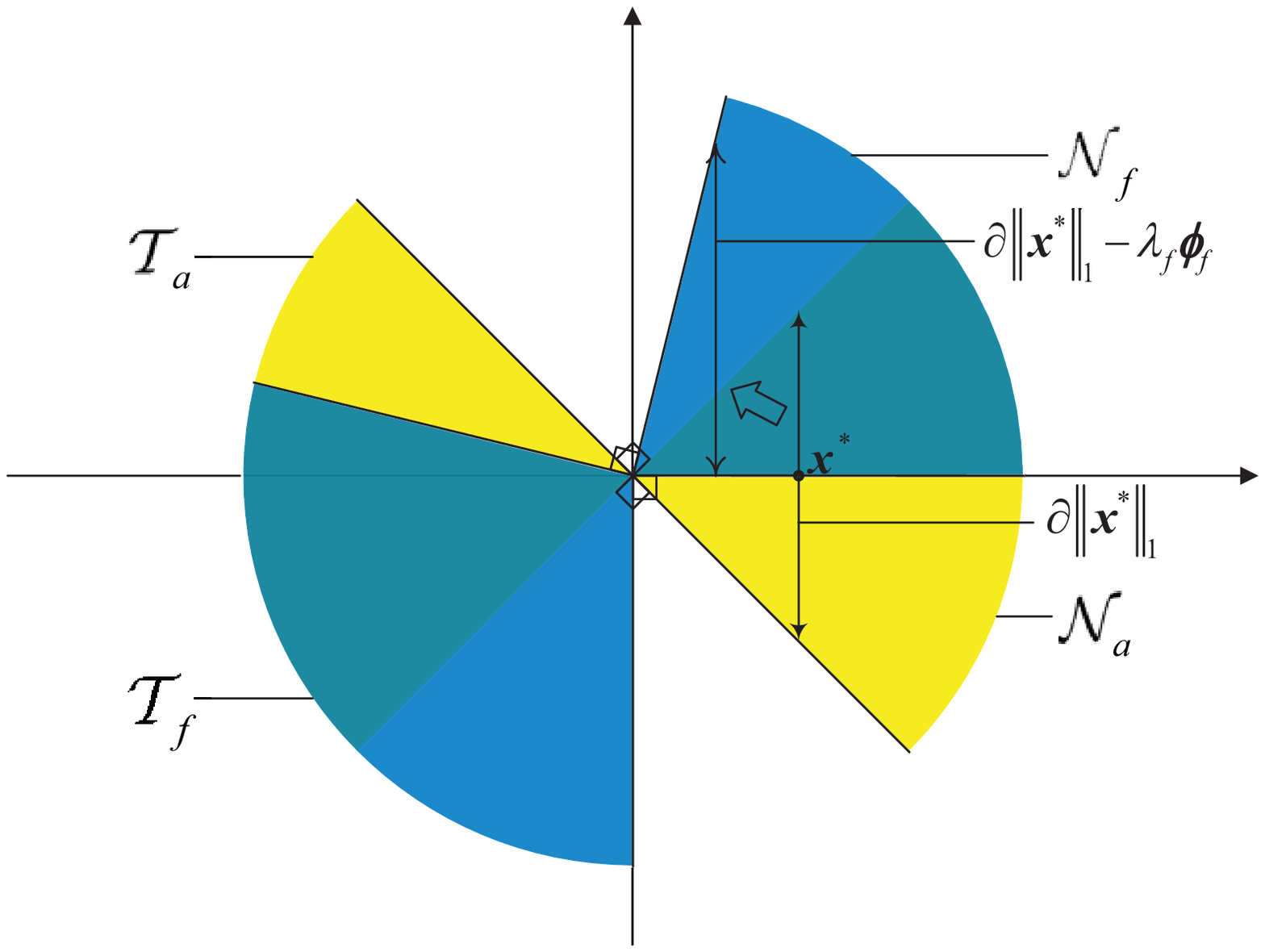}
		\label{fig_sixth_case}}
	\caption{The changes of normal cone and tangent cone after different shifts for sparse recovery. For reference, we draw the tangent cone $\mathcal{T}_a$ and normal come $\mathcal{N}_a$ in yellow for classical CS in all figures. In (b)-(f), the tangent cones $\mathcal{T}_\xi,~\xi=b,c,d,e,f$ and normal cones $\mathcal{N}_\xi,~\xi=b,c,d,e,f$ are shown in blue for the approach (\ref{eq: Main_Problem_Sparse}) under different shifts. Good shifts, such as (b) and (e), enlarge the normal cone and hence decrease the number of measurements. Bad shifts, such as (c), (d), and (f), narrow the normal cone and lead to a growth of sample size.}
	\label{fig: GeometricalAnalysisSparse}
\end{figure}

In what follows, we will illustrate these geometrical ideas by using two kinds of typical structured signals, namely, sparse vectors and low-rank matrices\footnote{The geometrical interpretation of block-sparse recovery is quite similar to that of sparse recovery, so we omit it here.}.

\subsection{Geometrical Interpretation of Procedure \eqref{eq: Main_Problem_Sparse}}
For the sparse recovery procedure \eqref{eq: Main_Problem_Sparse}, recall that $I$ is the support of $\vx^\star$. Since different shifts have different effects on the performance of \eqref{eq: Main_Problem_Sparse}, we define the following three kinds of shifts (or prior information), namely,
\begin{itemize}
\item a shift on $I$ if $\lambda \vphi$ satisfies
\begin{equation*}
    \left\{
    {\begin{array}{ll}
     \lambda \vphi_i \neq 0, & \exists~ i \in I,  \\
     \lambda \vphi_i   =  0, & \forall~ i \in I^c,
   \end{array} }
    \right.
\end{equation*}
\item a shift on $I^c$ if $\lambda \vphi$ satisfies
\begin{equation*}
    \left\{
    {\begin{array}{ll}
     \lambda \vphi_i   = 0, & \forall~  i \in I,  \\
     \lambda \vphi_i  \neq  0, & \exists~  i \in I^c,
   \end{array} }
    \right.
\end{equation*}
\item an arbitrary shift if $\lambda \vphi$ satisfies
\begin{equation*}
    \left\{
    {\begin{array}{ll}
     \lambda \vphi_i \neq  0, & \exists~  i \in I,  \\
     \lambda \vphi_i \neq  0, & \exists~  i \in I^c.
   \end{array} }
    \right.
\end{equation*}
\end{itemize}

For clarity, we consider a two dimensional example. We set $\vx^\star=[1,0]^T$ and consider different prior information: $\lambda_a\vphi_a=\[0,0\]^T,~\lambda_b\vphi_b=\[0.5,0\]^T~\lambda_c\vphi_c=\[-0.5,0\]^T,~\lambda_d\vphi_d=\[0,-1\]^T,~\lambda_e\vphi_e=\[0.5,-0.2\]^T,$ and $\lambda_f\vphi_f=\[0.5,-1\]^T$. Fig. \ref{fig: GeometricalAnalysisSparse} illustrates the changes in geometry under different shifts.

\begin{table}[!t]
\renewcommand{\arraystretch}{1.3}
\caption{$v_1$, $u_1$, upper bound I, upper bound II, and estimated Gaussian width under different shifts for sparse recovery.}
\label{table_1}
\centering
\begin{tabular}{c|c|c|c|c|c}
  \hline
  Shift      & $v_1$ & $u_1$& bound I & bound II & Result \\ \hline
  $\lambda_a \vphi_a$  & 2     & 1    & \textbf{1.68}   & 2             & 1.68 \\
  $\lambda_b \vphi_b$  & 1.25  & 0.25 & 1.49            & \textbf{1.25} & \textcolor[rgb]{0.00,0.00,1.00}{\textbf{1.25}} \\
  $\lambda_c \vphi_c$  & 3.25  & 2.25 & \textbf{1.80}   & 3.25          & 1.80 \\
  $\lambda_d \vphi_d$  & 5     & 2    & \textbf{1.87}   & 3             & 1.87 \\
  $\lambda_e \vphi_e$  & 1.69  & 0.29 & 1.62            & \textbf{1.29} & \textcolor[rgb]{0.00,0.00,1.00}{\textbf{1.29}} \\
  $\lambda_f \vphi_f$  & 4.25  & 1.25 & \textbf{1.85}   & 2.25          & 1.85 \\ \hline
\end{tabular}
\end{table}

We draw Fig. \ref{fig: GeometricalAnalysisSparse}(a) without shift for reference. The shifts in Fig. \ref{fig: GeometricalAnalysisSparse}(b) and \ref{fig: GeometricalAnalysisSparse}(c) are shifts on the support of $\vx^\star$. In Fig. \ref{fig: GeometricalAnalysisSparse}(b), the subdifferential moves toward the original, which enlarges the normal cone and hence decreases the number of measurements. The opposite result is shown in Fig. \ref{fig: GeometricalAnalysisSparse}(c). The shift in Fig. \ref{fig: GeometricalAnalysisSparse}(d) is a shift on the complement of the support of $\vx^\star$, which leads to a growth of sample size because the normal cone is narrowed in this case. The shifts in Fig. \ref{fig: GeometricalAnalysisSparse}(e) and \ref{fig: GeometricalAnalysisSparse}(f) are arbitrary shifts. We can know from Fig. \ref{fig: GeometricalAnalysisSparse}(e) and \ref{fig: GeometricalAnalysisSparse}(f) that different arbitrary shifts may lead to different results: the number of measurements of Fig. \ref{fig: GeometricalAnalysisSparse}(e) gets reduced  while that of Fig. \ref{fig: GeometricalAnalysisSparse}(f) gets increased.

\begin{figure}[H]
	\centering
	\subfloat[$\eta(\lambda_a\vPhi_a)=\[0,0,0\]^T$]{\includegraphics[width=2.3in]{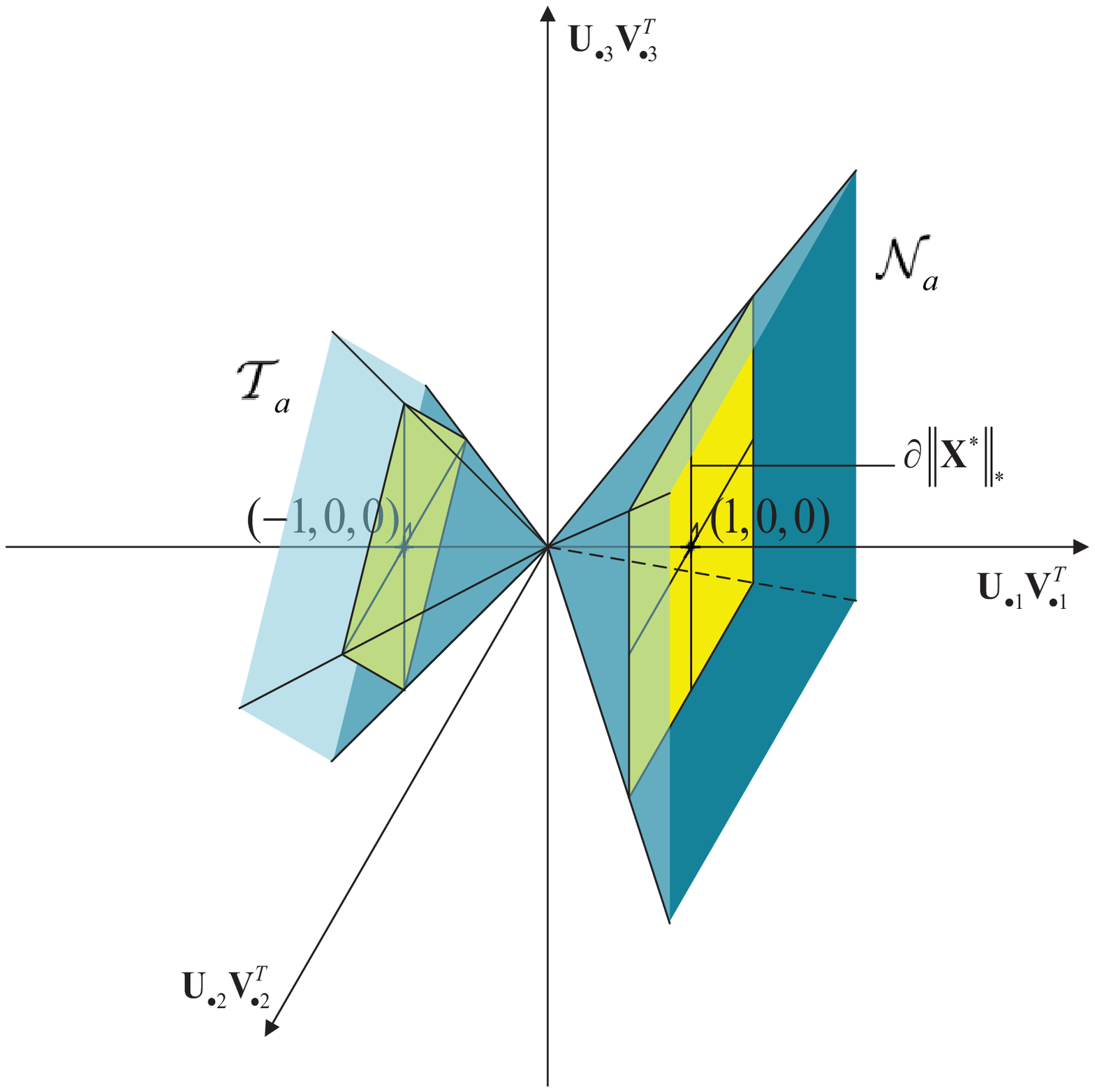}
		\label{fig_first_case_LowRank}}
	\hfil
	\subfloat[$\eta(\lambda_b\vPhi_b)=\[0.25,0,0\]^T$]{\includegraphics[width=2.3in]{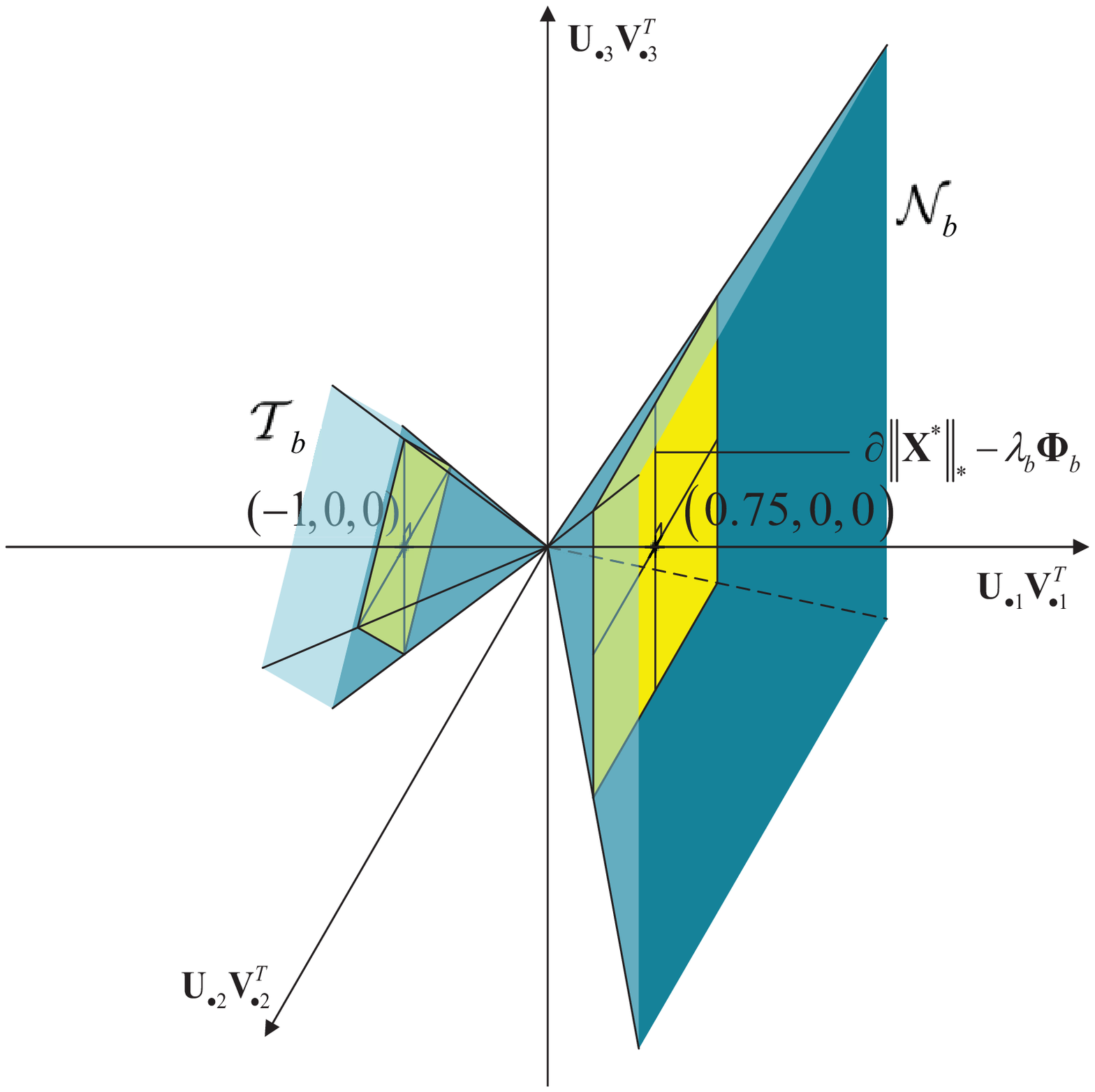}
		\label{fig_second_case_LowRank}}
	\hfil
	\subfloat[$\eta(\lambda_c\vPhi_c)=\[-0.25,0,0\]^T$]{\includegraphics[width=2.3in]{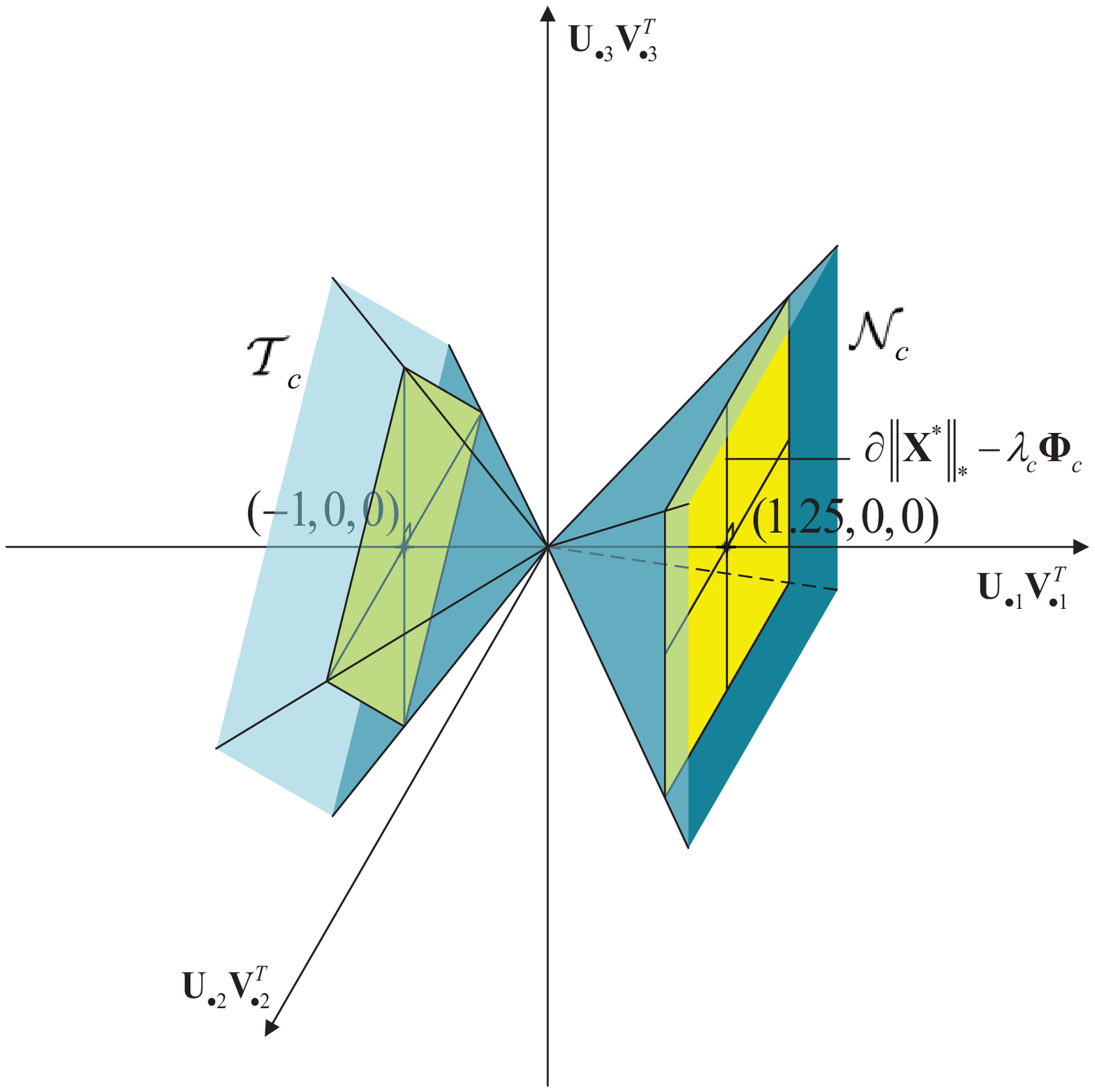}
		\label{fig_third_case_LowRank}}
	\hfil
	\subfloat[$\eta(\lambda_d\vPhi_d)=\[0,0,-0.5\]^T$]{\includegraphics[width=2.3in]{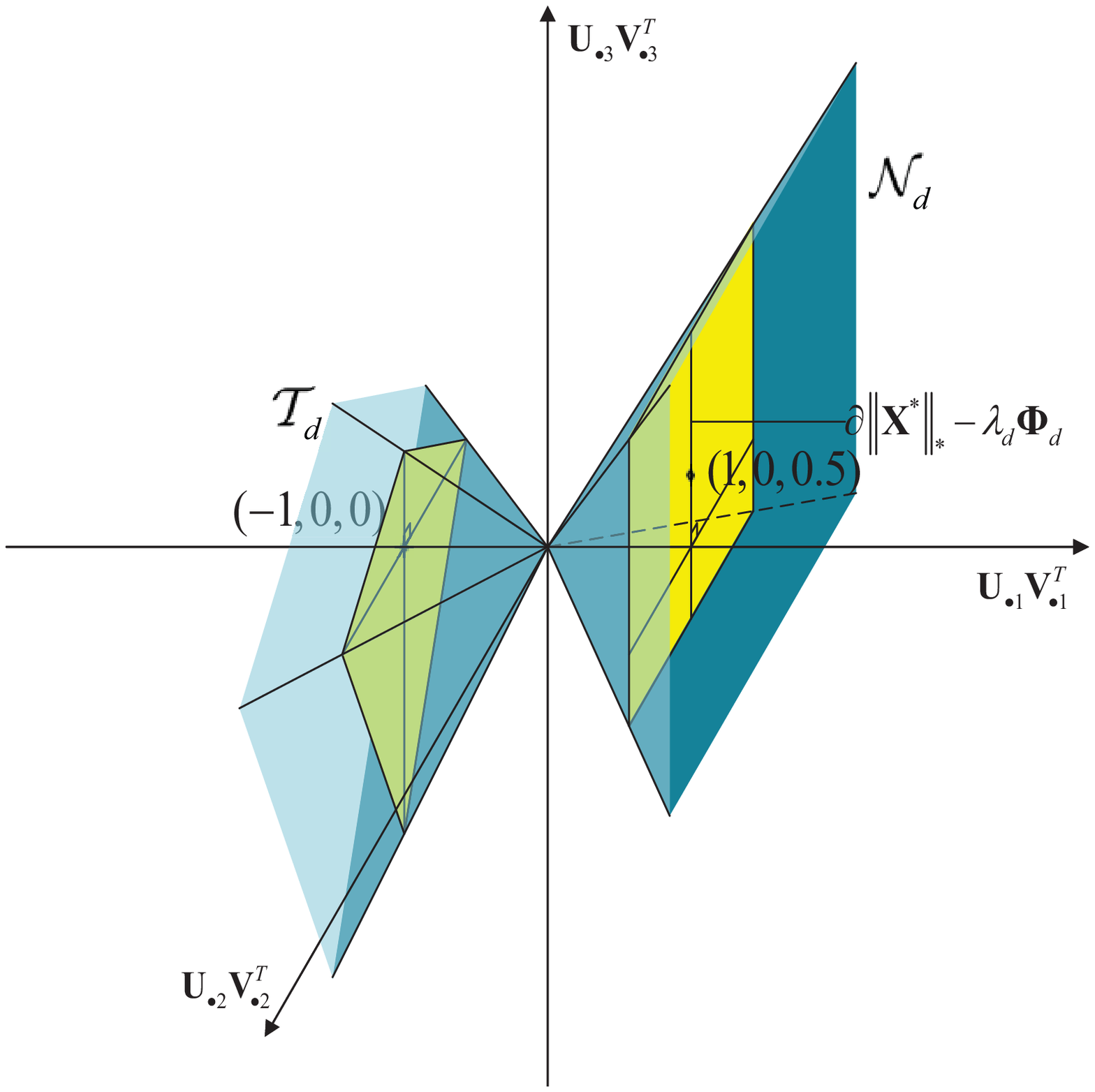}
		\label{fig_fourth_case_LowRank}}
	\hfil
	\subfloat[$\eta(\lambda_e\vPhi_e)=\[0.5,0,-0.2\]^T$]{\includegraphics[width=2.3in]{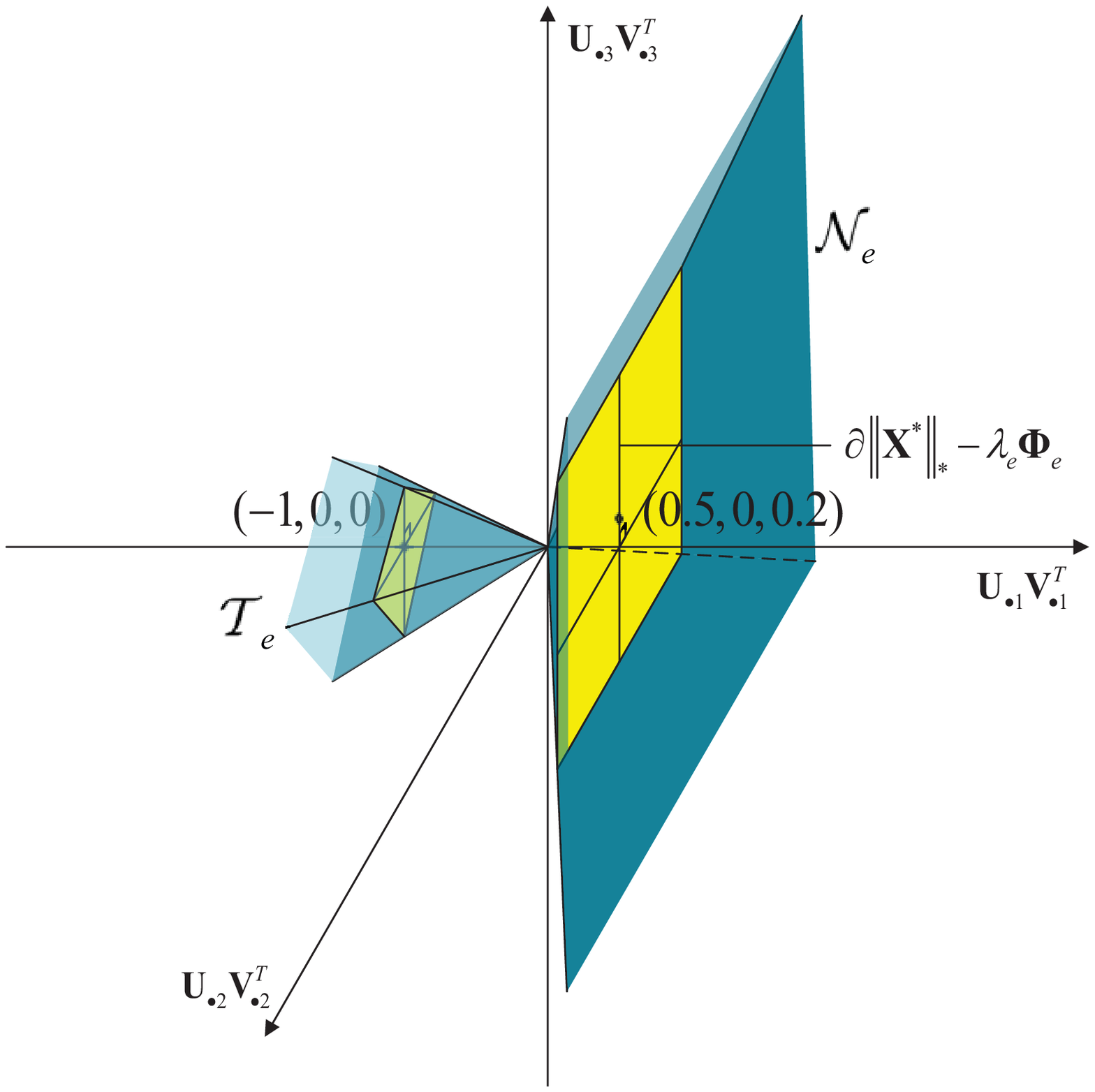}
		\label{fig_fifth_case_LowRank}}
	\hfil
	\subfloat[$\eta(\lambda_f\vPhi_f)=\[-0.25,0,-0.5\]^T$]{\includegraphics[width=2.3in]{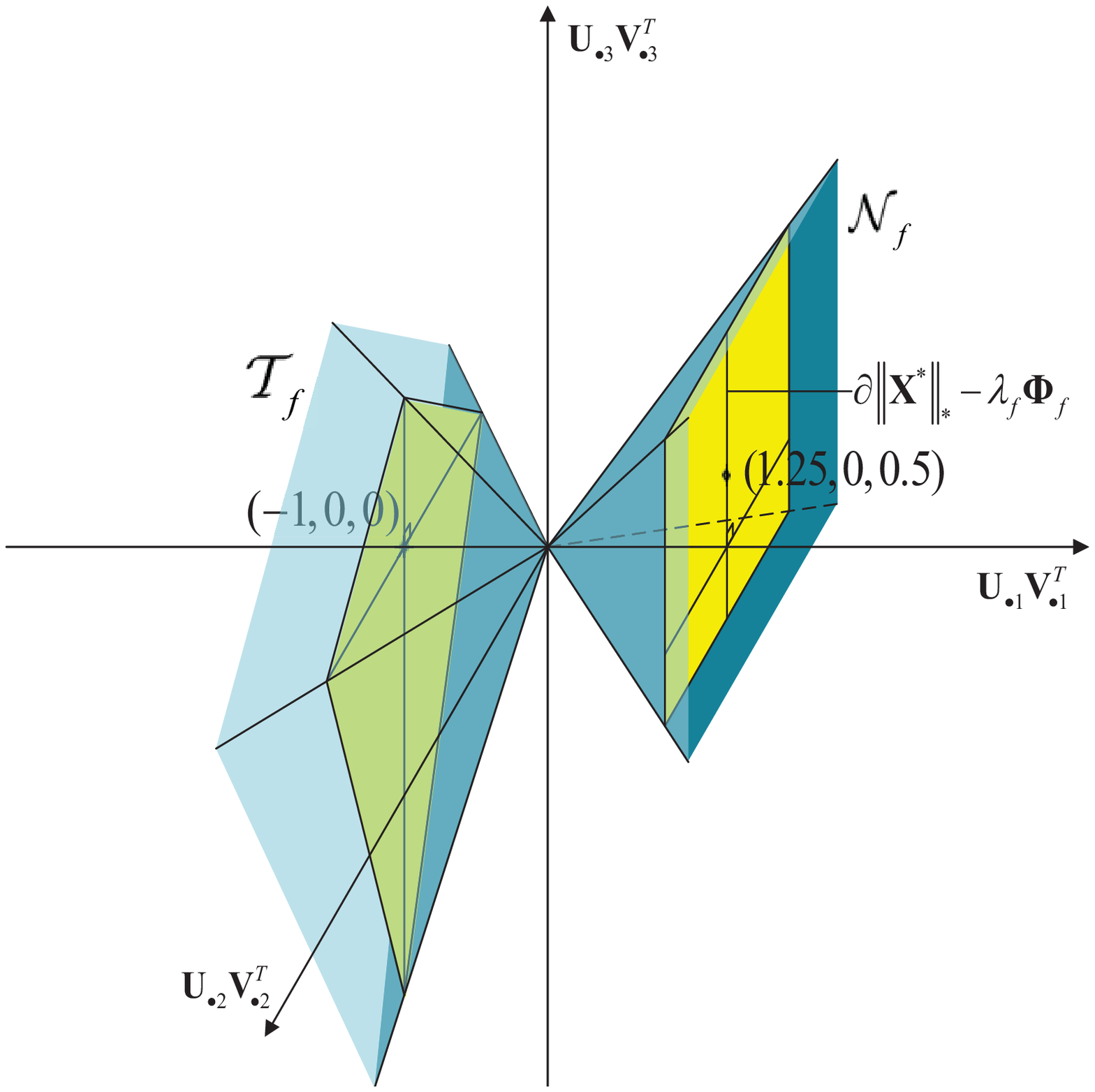}
		\label{fig_sixth_case_LowRank}}
	\caption{The changes of normal cone and tangent cone after different shifts for low-rank recovery.
		For reference, we draw the tangent cone $\mathcal{T}_a$ and normal come $\mathcal{N}_a$  for classical low-rank recovery in (a). In other subfigures, the tangent cones $\mathcal{T}_\xi,~\xi=b,c,d,e,f$ and normal cones $\mathcal{N}_\xi,~\xi=b,c,d,e,f$ are shown in blue for the approach (\ref{eq: Main_Problem_LowRank}) under different shifts. Good shifts, such as (b) and (e), enlarge the normal cone and hence decrease the number of measurements. Bad shifts, such as (c), (d), and (f), narrow the normal cone and lead to a growth of sample size.  }
	\label{fig: GeometricalAnalysisLowRank}
\end{figure}

Table \ref{table_1} lists related parameters ($v_1$ and $u_1$) and estimated Gaussian widths under different shifts. In each case, we calculate the two upper bounds in Lemma \ref{lm: GaussianWidth_Sparse} and obtain the estimated Gaussian width by minimizing them. The results illustrate that the shifts $\lambda_b \vphi_b$ and $\lambda_e \vphi_e$ outperform the no shift case while the other shifts present an opposite result.

Therefore, we might conclude that with proper shifts, our approach can outperform BP.

\subsection{Geometrical Interpretation of Procedure \eqref{eq: Main_Problem_LowRank}}

For low-rank matrix recovery procedure \eqref{eq: Main_Problem_LowRank}, recall that $\mathcal{S}$ is the space of matrices in the column or row space of $\vX^\star$. We similarly consider three different kinds of shifts (or prior information), namely,
\begin{itemize}
\item a shift on $\mathcal{S}$ if $ \vPhi$ satisfies
\begin{equation*}
    \left\{
    {\begin{array}{ll}
     \mathcal{P}_\mathcal{S} (\vPhi) &\neq \bm{0}, \\
     \mathcal{P}_{\mathcal{S}^{\bot}}(\vPhi)   &=  \bm{0},
   \end{array} }
    \right.
\end{equation*}
\item a shift on $\mathcal{S}^{\bot}$ if $ \vPhi$ satisfies
\begin{equation*}
    \left\{
    {\begin{array}{ll}
    \mathcal{P}_\mathcal{S} (\vPhi) &= \bm{0}, \\
    \mathcal{P}_{\mathcal{S}^{\bot}}(\vPhi)  &\neq   \bm{0},
   \end{array} }
    \right.
\end{equation*}
\item an arbitrary shift if $\vPhi$ satisfies
\begin{equation*}
    \left\{
    {\begin{array}{ll}
     \mathcal{P}_\mathcal{S} (\vPhi) &\neq \bm{0}, \\
     \mathcal{P}_{\mathcal{S}^{\bot}}(\vPhi) &\neq  \bm{0}.
   \end{array} }
    \right.
\end{equation*}
\end{itemize}

For clarity, we consider a three dimensional example. Let
\begin{equation*}
  \vX^\star=\vU\left[
                              \begin{array}{ccc}
                                1 & 0 & 0 \\
                                0 & 0 & 0 \\
                                0 & 0 & 0 \\
                              \end{array}
                            \right]
                            \vV^T,
\end{equation*}
be the SVD of $\vX^\star$, and let $\vPhi=\sum_{k=1}^{3}[\eta(\vPhi)]_k \vU_{\cdot k} \vV_{\cdot k}^T$ with $\eta(\vPhi) \in \R^3$. We consider different prior information: $\eta(\lambda_a\vPhi_a)=\[0,0,0\]^T,~ \eta(\lambda_b\vPhi_b)=\[0.25,0,0\]^T,~\eta(\lambda_c\vPhi_c)=\[-0.25,0,0\]^T,~\eta(\lambda_d\vPhi_d)=\[0,0,-0.5\]^T,
~\eta(\lambda_e \vPhi_e)=[0.5,0,-0.2]^T$, and $\eta(\lambda_f\vPhi_f)=\[-0.25,0,-0.5\]^T$. It is not hard to check that $\lambda_b\vPhi_b$ and $\lambda_c\vPhi_c$ are shifts on $\mathcal{S}$, $\lambda_d\vPhi_d$ is a shift on $\mathcal{S}^{\perp}$, and $\lambda_e \vPhi_e$ and $\lambda_f \vPhi_f$ are arbitrary shifts. The changes in geometry under different shifts are shown in Fig. \ref{fig: GeometricalAnalysisLowRank}.

For reference, we draw Fig. \ref{fig: GeometricalAnalysisLowRank}(a) for the standard low-rank recovery procedure. Compared with Fig. \ref{fig: GeometricalAnalysisLowRank}(a), the shifts in Fig. \ref{fig: GeometricalAnalysisLowRank}(b) and \ref{fig: GeometricalAnalysisLowRank}(e) move toward the original, which enlarges the normal cone and hence decreases the number of measurements. On the contrary, the shifts in Fig. \ref{fig: GeometricalAnalysisLowRank}(c), \ref{fig: GeometricalAnalysisLowRank}(d), and \ref{fig: GeometricalAnalysisLowRank}(f) narrow the normal cone and hence lead to a growth of the sample size.

The related parameters ($v_3$ and $u_3$) and estimated Gaussian widths in Lemma \ref{lm: GaussianWidth_Blocksparse} are shown in Table \ref{table_2}. The results illustrate that the shifts $\lambda_b \vPhi_b$ and $\lambda_e \vPhi_e$ outperform the no shift case while the other shifts present an opposite result. Thus we may similarly conclude that with proper prior information, the procedure \eqref{eq: Main_Problem_LowRank} can improve the performance of the standard recovery method.

\begin{table}[!t]
\renewcommand{\arraystretch}{1.3}
\caption{$v_3$, $u_3$, upper bound I, upper bound II, and estimated Gaussian width under different shifts for low-rank matrix recovery.}
\label{table_2}
\centering
\begin{tabular}{c|c|c|c|c|c}
  \hline
  Shift               & $v_3$   & $u_3$  & bound I    & bound II & Result \\ \hline
  $\lambda_a\vPhi_a$  & 3       & 1      & \textbf{8.9707} & 15           & 8.9707   \\
  $\lambda_b\vPhi_b$  & 2.5625  & 0.5625 & \textbf{8.9657} & 10.625       & \textcolor[rgb]{0.00,0.00,1.00}{\textbf{8.9657}}  \\
  $\lambda_c\vPhi_c$  & 3.5625  & 1.5625 & \textbf{8.9754} & 20.625       & 8.9754 \\
  $\lambda_d\vPhi_d$  & 4.25    & 1.25   & \textbf{8.9793} & 17.5         & 8.9793 \\
  $\lambda_e\vPhi_e$  & 2.69    & 0.29   & 8.9674          & \textbf{7.9} & \textcolor[rgb]{0.00,0.00,1.00}{\textbf{7.9}} \\
  $\lambda_f\vPhi_f$  & 4.8125  & 1.8125 & \textbf{8.9818} & 23.125       & 8.9818 \\\hline
\end{tabular}
\end{table}

\subsection{Guidance for Improving Prior Information}\label{Guidance}
The proofs of Theorems \ref{thm: SparseCase}-\ref{thm: LowRankCase} have revealed that $v_{i}$ and $u_{i}$ ($i = 1, 2, 3$) play a key role in evaluating the performance of our recovery procedures. More precisely, the less $v_i$ and $u_i$, the better the recovery performance. Therefore, these parameters might provide practical guidance to improve the quality of prior information.

\textbf{Sparse recovery.} In this case,
\isdraft{  
\begin{equation}
v_1= \sum \limits_{i \in I} (\sign(x_i^\star)-\lambda \vphi_i)^2 + \sum \limits_{i \in I^c} (1+|\lambda \vphi_i|)^2
\end{equation}
}          
{
\begin{equation}
    v_1= \sum \limits_{i \in I} (\sign(\vx_i^\star)-\lambda \vphi_i)^2
     + \sum \limits_{i \in I^c} (1+|\lambda \vphi_i|)^2
\end{equation}
}
and
\begin{align}
u_1= \sum \limits_{i \in I} (\sign(\vx_i^\star)-\lambda \vphi_i)^2 + \sum \limits_{i \in I^c} (\lambda \vphi_i)^2.
\end{align}
It is not hard to find that
\begin{itemize}
  \item For $i \in I^c$, $\vphi_i=0$  is preferred since $\sum_{i \in I^c} (1+|\lambda \vphi_i|)^2$ and $\sum_{i \in I^c} (\lambda \vphi_i)^2$ cannot be less than that in the no shift case.
  \item For $i \in I$, $\sum_{i \in I} (\sign(\vx_i^\star)-\lambda \vphi_i)^2$ may be smaller than that in the no shift case, provided that prior information is suitable, for example,
  $\lambda \vphi_i = \sign(\vx_i^\star)/2$ for $i \in I$.
\end{itemize}
These observations suggest the following strategies to reduce $v_1$ and $u_1$ and hence to improve the performance of \eqref{eq: Main_Problem_Sparse}
\begin{itemize}
  \item[(S1)] If the sparsity level $s$ of $\vx^{\star}$ is known, then we can keep only top $s$ principal components of $\vphi$ (in terms of absolute values) and adjust the new shift as follows
      $$
        \tilde{\lambda} \tilde{\vphi}= \kappa \cdot \sign(\vphi_T),
      $$
      where $T$ is the support of $s$ principal components of $\vphi$ and $\kappa \in (0,1)$ is an adjustable parameter.
  \item[(S2)] If the sparsity level $s$ of $\vx^{\star}$ is unknown, then we might estimate the stable sparsity level first \cite{lopes2013estimating}$$ s=\frac{\norm{\vphi}_1^2}{\norm{\vphi}_2^2}.$$
       Thus we can repeat the procedures in (S1) and obtain the new shift.
\end{itemize}

\textbf{Block-sparse recovery.} In this case, standard calculation leads to
\isdraft{               
\begin{equation}
v_2 =\sum \limits_{b \in \mathcal{B}} \norm{\frac{\vx_{V_b}^\star}{\norm{\vx_{V_b}^\star}_2}-\lambda \vphi_{V_b}}_2^2\\+\sum \limits_{b \notin \mathcal{B}} \(1+\lambda \norm{\vphi_{V_b}}_2\)^2
\end{equation}
}{                      
\begin{align*}
v_2 =\sum \limits_{b \in \mathcal{B}} \norm{\frac{\vx_{V_b}^\star}{\norm{\vx_{V_b}^\star}_2}-\lambda \vphi_{V_b}}_2^2 + \sum \limits_{b \notin \mathcal{B}} \(1+\lambda \norm{\vphi_{V_b}}_2\)^2
\end{align*}
}
and
\begin{equation}
u_2 =\sum \limits_{b \in \mathcal{B}} \norm{\frac{\vx_{V_b}^\star}{\norm{\vx_{V_b}^\star}_2}-\lambda \vphi_{V_b}}_2^2
+\sum \limits_{b \notin \mathcal{B}}\norm{\lambda \vphi_{V_b}}_2^2.
\end{equation}

Similarly, we can find that
\begin{itemize}
  \item For $b \notin \mathcal{B}$, $\vphi_{V_b}=0$ is preferred because $\sum_{b \notin \mathcal{B}} \(1+\lambda \norm{\vphi_{V_b}}_2\)^2$ and $\sum_{b \notin \mathcal{B}}\norm{\lambda \vphi_{V_b}}_2^2$ cannot be less than that in the no shift case.
  \item For $b \in \mathcal{B}$, $\sum_{b \in \mathcal{B}} \|{\vx_{V_b}^\star}/{\|\vx_{V_b}^\star\|_2}-\lambda \vphi_{V_b}\|_2^2$ may be less than that in the no shift case, provided that prior information is proper, for example, $\lambda\vphi = \frac{1}{2} \sum_{b \in \mathcal{B} }{\vx_{V_b}^\star}/\norm{\vx_{V_b}^\star}_2$.
\end{itemize}
Thus we can reduce $v_2$ and $u_2$ by the following strategies
\begin{itemize}
  \item[(B1)] If the number of non-zero blocks $l$ of $\vx^{\star}$ is known, then we can keep only top $l$ principal blocks of $\vphi$ (in terms of the $\ell_2$-norm of the blocks) and obtain the new shift
      \begin{equation*}
        \tilde{\lambda} \tilde{\vphi} = \kappa \sum_{b \in \tilde{\mathcal{B}}}\frac{\vx_{V_b}^\star}{\norm{\vx_{V_b}^\star}_2},
      \end{equation*}
  where $\tilde{\mathcal{B}}$ is the support of principal blocks of $\vphi$ and $\kappa \in (0, 1)$.
  \item[(B2)] If the number of non-zero blocks $l$ of $\vx^{\star}$ is unknown, then we estimate the stable number of blocks as in \cite{lopes2013estimating}
  $$ l=\frac{\norm{\vphi}_{2,1}^2}{\norm{\vphi}_2^2}.$$
  Thus, the new shift is gotten by repeating the procedures in (B1).
\end{itemize}

\textbf{Low-rank recovery.} In the low-rank matrix recovery case, direct calculation yields
\isdraft{               
\begin{equation} \label{parameter_v_3}
v_3=\norm{\vU\vV^T -\lambda \mathcal{P}_{\mathcal{S}}(\vPhi)}_F^2 +\max \limits_{\vW \in \mathcal{S}^\bot, \norm{\vW} \le 1} \norm{\vW-\lambda \mathcal{P}_{\mathcal{S}^\bot}(\vPhi)}_F^2,
\end{equation}
}{                      
\begin{multline} \label{parameter_v_3}
    v_3=\norm{\vU\vV^T -\lambda \mathcal{P}_{\mathcal{S}}(\vPhi)}_F^2 \\+\max \limits_{\vW \in \mathcal{S}^\bot, \norm{\vW} \le 1} \norm{\vW-\lambda \mathcal{P}_{\mathcal{S}^\bot}(\vPhi)}_F^2,
\end{multline}}
and
\begin{equation}\label{parameter_u_3}
  u_3=\norm{\vU\vV^T -\lambda \mathcal{P}_{\mathcal{S}}(\vPhi)}_F^2+\norm{\lambda \mathcal{P}_{\mathcal{S}^\bot}(\vPhi)}_F^2.
\end{equation}
From the above expressions, it is not hard to see that the left and right singular vectors of non-zero singular values of $\vPhi$ play a critical role in improving the quantity of priori information. Let $\vPhi= \hat{\vU} \hat{\vSigma} \hat{\vV}$ be the SVD of $\vPhi$, then we can similarly obtain the following strategies
\begin{itemize}
  \item[(L1)]  If the rank $r$ of $\vX^{\star}$ is known,  then we improve the new shift as follows
      $$
        \tilde{\lambda} \tilde{\vPhi}= \kappa \cdot \hat{\vU}_r\hat{\vV}_r^T,
      $$
      where $\hat{\vU}_r$ and $\hat{\vV}_r$ consist of the first $r$ left and right singular vectors of $\vPhi$ respectively and $\kappa \in (0,1)$.
 \item[(L2)] If the rank $r$ of $\vX^{\star}$ is unknown, we should estimate the numerical rank first. For example, we can use the way in \cite{rudelson2007sampling}
  $$ r=\frac{\norm{\vPhi}_{\text{F}}^2}{\norm{\vPhi}^2}.$$
  Repeating the procedures in (L1) completes the new shift.
\end{itemize}

%

\section{Numerical Simulations}  \label{sec: Simulation}


In this section, we carry out some numerical simulations to verify the correctness of our theoretical results.

\subsection{Phase Transition for Sparse Recovery}
In this experiment, we draw some phase transition curves in the absence of noise for different kinds of prior information: no shift, shifts on the support, shifts on the complement of the support, and arbitrary shifts. The original signal $\vx^\star \in \R^n$ is an $s$-sparse random vector whose nonzero entries are drawn from the standard Gaussian distribution and the measurement matrix $\vA \in \R^{m \times n}$ is a Bernoulli matrix with i.i.d. entries obeying the symmetric Bernoulli distribution. We set $n=128$ and $tol=10^{-2}$ for all the experiments. For a particular pair of $s$ and $m$, we make 50 trials, count the number of trials which succeed to recover $\vx^\star$, and calculate the related probability. Let $\hat{\vx}$ be the solution of \eqref{eq: Main_Problem_Sparse}. If the solution of a trial satisfies
$$
\frac{\norm{\vx^\star - \hat{\vx}}_2}{\norm{\vx^\star}_2} < tol,
$$
then we claim it as a successful trial. We increase both $m$ and $s$ from $0$ to $n$ with step $2$, then we can obtain a phase transition curve.

We consider six different shifts:
\begin{enumerate}
  \item[(a)] $\lambda \vphi=\bm{0}$. This is the classical CS model.
  \item[(b)] $\lambda \vphi= \text{sign}(\vx^\star)/2$. This is a shift on the support.
  \item[(c)] $\lambda \vphi=-\text{sign}(\vx^\star)/2$. This is a shift on the support.
  \item[(d)] $\lambda \vphi_{i}=0$ for $i \in I$ and $\lambda \vphi_{i}=1$ for $i \in I^c$. This is a shift on the complement of the support.
  \item[(e)] $\lambda \vphi_{i}=\text{sign}(\vx^\star_i)/2$ for $i \in I$ and $\lambda \vphi_{i}=0$ for $i \in I^c$ except an arbitrary $i \in I^c$ satisfying $\lambda \vphi_i=1/4$. This is an arbitrary shift.
  \item[(f)] $\lambda \vphi_{i}=-\text{sign}(\vx^\star_i)/2$ for $i \in I$ and $\lambda \vphi_{i}=1$ for $i \in I^c$. This is an arbitrary shift.
\end{enumerate}

\begin{figure}[H]
	\centering
	\subfloat[$\lambda \vphi=\bm{0}$]{\includegraphics[width=2.3in]{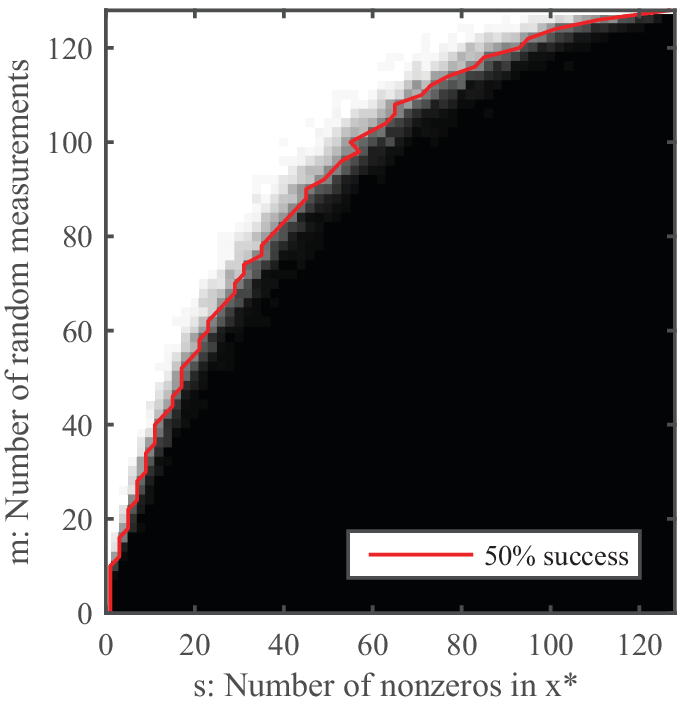}
		\label{fig2_first_case}}
	\hfil
	\subfloat[$\lambda \vphi= \frac{\text{sign}(\vx^\star)}{2}$]{\includegraphics[width=2.3in]{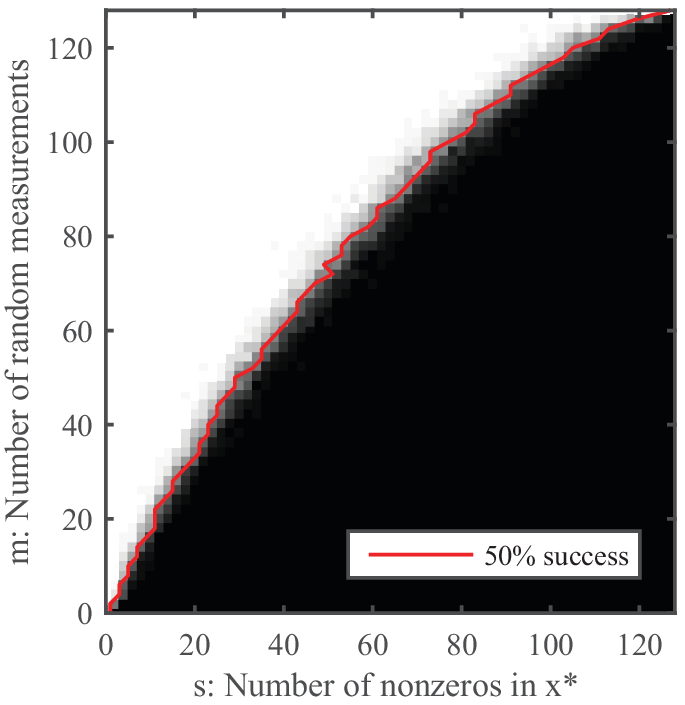}
		\label{fig2_second_case}}
	\hfil
	\subfloat[$\lambda \vphi=-\frac{\text{sign}(\vx^\star)}{2}$]{\includegraphics[width=2.3in]{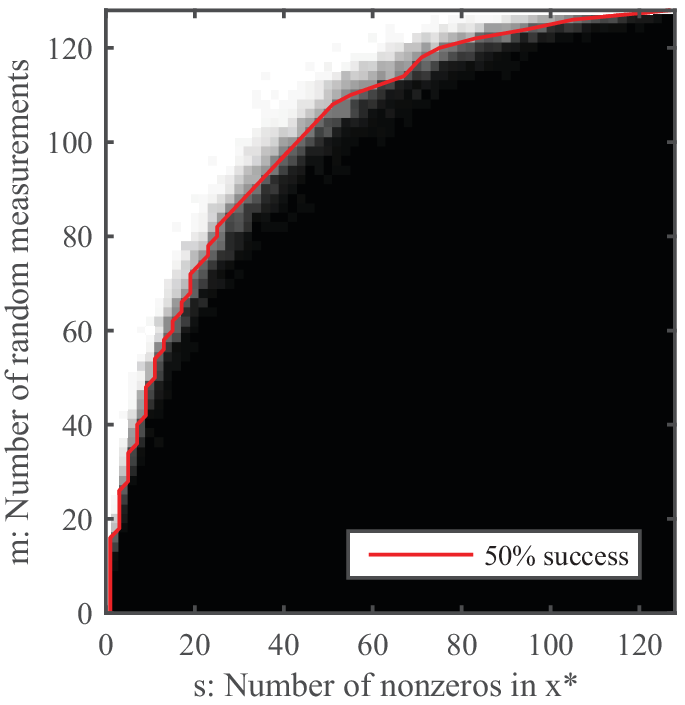}
		\label{fig2_third_case}}
	\hfil
	\subfloat[$\lambda \vphi_{i}=0$ for $i \in I$ and $\lambda \vphi_{i}=1$ for $i \in I^c$]{\includegraphics[width=2.3in]{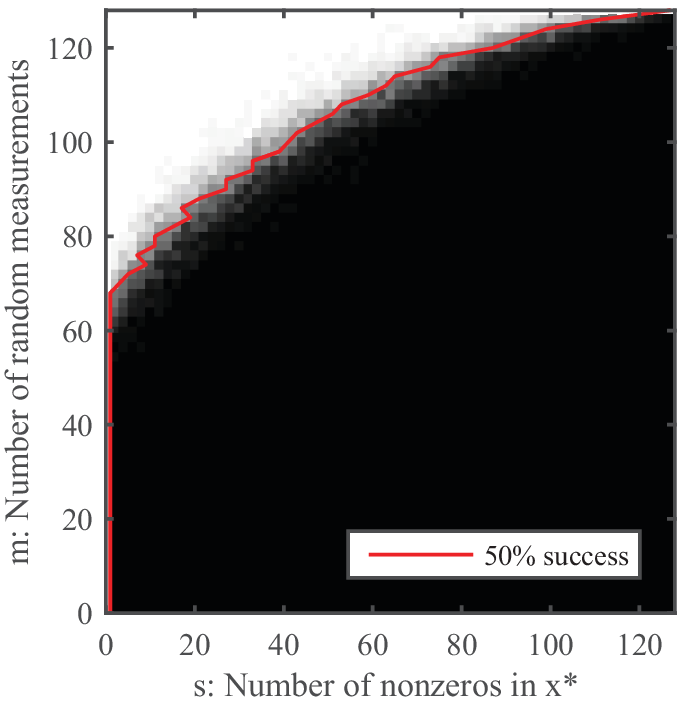}
		\label{fig2_fourth_case}}
	\hfil
	\subfloat[$\lambda \vphi_{i}=\text{sign}(\vx^\star_i)/2$ for $i \in I$ and $\lambda \vphi_{i}=0$ for $i \in I^c$ except an arbitrary $i \in I^c$ satisfying $\lambda \vphi_i=1/4$]{\includegraphics[width=2.3in]{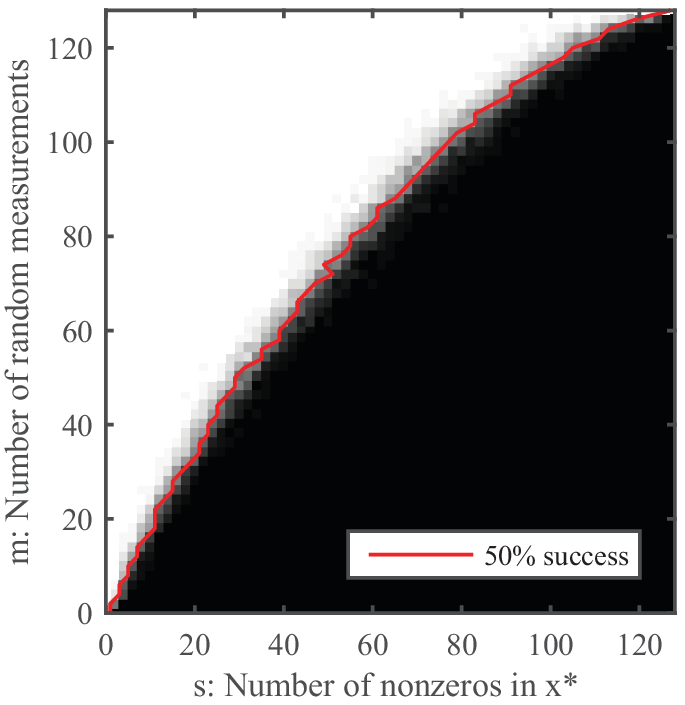}
		\label{fig2_fifth_case}}
	\hfil
	\subfloat[ $\lambda \vphi_{i}=-\text{sign}(\vx^\star_i)/2$ for $i \in I$ and $\lambda \vphi_{i}=1$ for $i \in I^c$]{\includegraphics[width=2.3in]{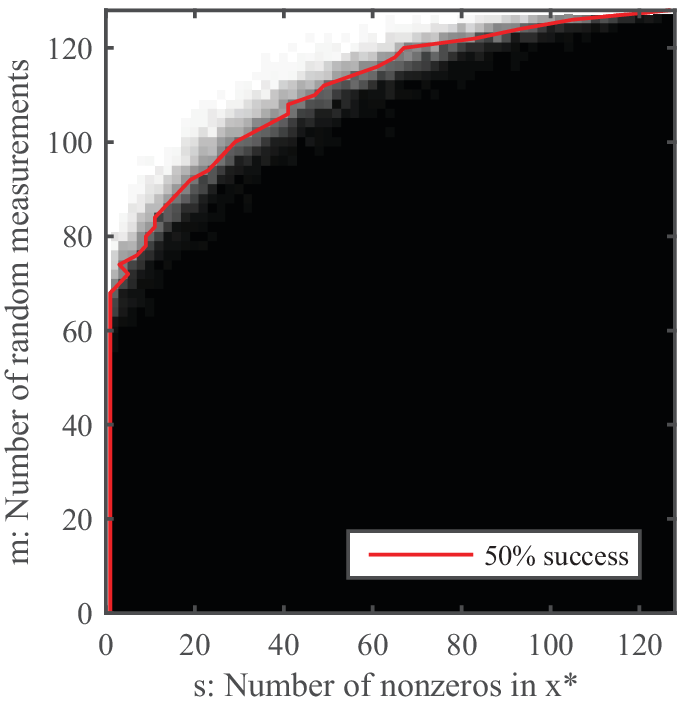}
		\label{fig2_sixth_case}}
	\caption{The phase transition curves for different shifts. The brightness of each point reflects the observed probability of success, ranging from black (0\%) to white (100\%).}
	\label{fig: PhaseTransition}
\end{figure}

The results are shown in Fig \ref{fig: PhaseTransition}. For shifts on the support, Fig. \ref{fig: PhaseTransition}(b) shows an improved performance while Fig. \ref{fig: PhaseTransition}(c) presents a deteriorative performance in contrast to the standard CS result in Fig. \ref{fig: PhaseTransition}(a). Comparing Fig. \ref{fig: PhaseTransition}(d) with Fig. \ref{fig: PhaseTransition}(a), we realize that the shift on the complement of the support makes the number of measurements increase whatever the sparsity level is. In Fig. \ref{fig: PhaseTransition}(e), the simulation result shows that this arbitrary shift improves the performance a lot compared with Fig. \ref{fig: PhaseTransition}(a). However, Fig. \ref{fig: PhaseTransition}(f) presents an opposite result for another arbitrary shift. All of these numerical results are consistent with our theoretical result (Theorem \ref{thm: SparseCase}) and geometrical interpretation. From these results, we can conclude that if prior information is good enough, then our method can perform better than BP.


\subsection{Performance Comparisons with Other Approaches for Sparse Recovery}

We next run some experiments to compare the performance of BP,  $\ell_1$-$\ell_1$ minimization, $\ell_1$-$\ell_2$ minimization, the proposed method (denoted by maximizing correlation (MC)), improved MC with known sparsity, and improved MC with unknown sparsity. Here, two improved MCs follow the strategies in Section \ref{Guidance} for sparse recovery.

\begin{figure}[H]
	\centering
	\subfloat[$\vz$: random sparse perturbation]{\includegraphics[width=3.5in]{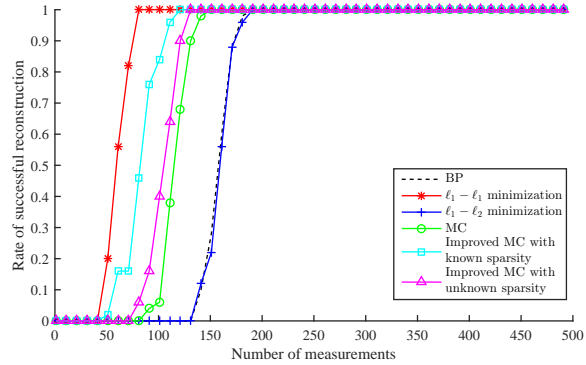}
		\label{fig6_first_case}}
	\hfil
	\subfloat[$\vz$: random dense perturbation]{\includegraphics[width=3.5in]{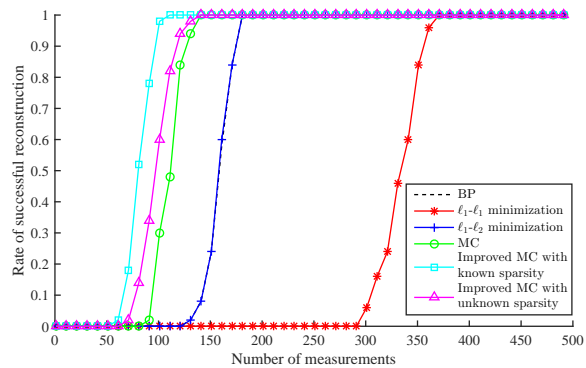}
		\label{fig6_second_case}}
	\caption{Rate of successful reconstruction v.s. number of measurements under prior information $\vphi=\vx^\star+\vz$ for BP, $\ell_1$-$\ell_1$ minimization, $\ell_1$-$\ell_2$ minimization, maximizing correlation (MC), improved MC with known sparsity, and improved MC with unknown sparsity. (a) $\vz$ is a random sparse perturbation. (b) $\vz$ is a random dense perturbation.}
	\label{fig: PerformanceComparison}
\end{figure}
 The original signal $\vx^\star$ is a sparse vector with dimension $n=500$ and sparsity $s=50$. The nonzero entries of $\vx^\star$ are drawn from the standard Gaussian distribution. The prior information is generated as $\vphi=\vx^\star+\vz$. We consider two kinds of perturbation vector $\vz$:
\begin{itemize}
	\item 	Random sparse perturbation. $\vz$ is a $20$-sparse vector whose nonzero entries are drawn from a zero-mean Gaussian random variable with standard deviation $0.5$. The supports of $\vx^\star$ and $\vz$ coincide in $16$ positions and differ in $4$ positions.
	\item Random dense perturbation. The entries of $\vz$ are drawn from a zero-mean Gaussian variable with standard deviation $0.1$.
\end{itemize}
The measurement matrix $\vA \in \R^{m \times n}$ is a Bernoulli matrix. The observation noise $\vn \in \R^m$ is a random vector whose entries are independently drawn from a zero-mean Gaussian distribution with standard deviation $0.01$. We set the upper bound of the noise vector $\delta=\norm{\vn}_2$. For a fixed number of measurements $m$, we make $50$ trials, and calculate the successful probability as before. We set $\lambda=1$ for $\ell_1$-$\ell_1$ minimization, $\ell_1$-$\ell_2$ minimization and MC, and set $\kappa=0.95$ for two improved MCs.

The results under random sparse perturbation are shown in Fig. \ref{fig: PerformanceComparison}(a).  The figure shows that $\ell_1$-$\ell_1$ minimization performs the best. This is because $\vphi - \vx^{\star}$ is a sparse vector in this case and the $\ell_1$-norm is effective to promote sparsity. We also find that all three MCs outperform BP and $\ell_1$-$\ell_2$ minimization. In addition, the results indicates that the strategies for improving prior information are effective since the two improved MCs show a better performance than the normal one. The plot also illustrates that the improved MC with known sparsity outperforms the one with unknown sparsity.

Fig. \ref{fig: PerformanceComparison}(b) shows the experiment results under random dense perturbation. The plot shows that all three MCs perform better than the other methods. $\ell_1$-$\ell_1$ minimization performs the worst in this case since $\vphi - \vx^{\star}$ is not a sparse vector anymore. BP and $\ell_1$-$\ell_2$ minimization almost coincide everywhere. For three MCs, a similar phenomenon is observed that the two improved MCs have a better performance than the normal one and that the improved MC with known sparsity outperforms the one with unknown sparsity.

In conclusion, whether the perturbation is sparse or dense, the proposed method can have a relatively good performance. Furthermore, the simulations validate the effectiveness of  the proposed guidance for improving prior information.

\subsection{Phase Transition for Low-rank Recovery}
For low-rank matrix recovery, we also draw some phase transition curves in the absence of noise for different kinds of prior information: no shift, shifts on $\mathcal{S}$, shifts on $\mathcal{S}^{\perp}$, and arbitrary shifts. Let $\vX \in \R^{n \times n}$ be a Gaussian matrix with i.i.d. entries satisfying the standard Gaussian distribution. Let $\vX = \tilde{\vU} \tilde{\vSigma} \tilde{\vV}^T = \sum_{k =1}^{n} \tilde{\sigma}_k \tilde{\vU}_{\cdot k} \tilde{\vV}_{\cdot k}^T$ be the SVD of $\vX$. The original signal $\vX^\star=\sum_{k =1}^{r} \tilde{\sigma}_k \tilde{\vU}_{\cdot k} \tilde{\vV}_{\cdot k}^T \in \R^{n \times n}$ is a rank-$r$ matrix
and the measurement matrices $\{\vA^j\}_{j=1}^{m}$ are independent Bernoulli matrices. We set $n=32$ and $tol=10^{-2}$ for all the experiments. For a particular pair of $r$ and $m$, we make 50 trials, count the number of trials which succeed to recover $\vX^\star$, and calculate the related probability. If the solution of a trial satisfies
$$
\frac{\norm{\vX^\star - \hat{\vX}}_\text{F}}{\norm{\vX^\star}_\text{F}} < tol,
$$
we claim it as a successful trial. Let $r$ increase from $0$ to $n$ with step 1 and $m$ increase from $0$ to $n^2$ with step $n$, then we can obtain a phase transition curve.

We consider six different shifts:
\begin{enumerate}
  \item[(a)] $\lambda \vPhi=\bm{0}$. This is the classical CS model for low-rank recovery.
  \item[(b)] $\lambda \vPhi= \vU\vV^T/2$. This is a shift on $\mathcal{S}$.
  \item[(c)] $\lambda \vPhi=-\vU\vV^T/2$. This is a shift on $\mathcal{S}$.
  \item[(d)] $\lambda \vPhi=\vU'\vV'^T/2$. This is a shift on $\mathcal{S}^{\perp}$.
  \item[(e)] $\lambda \vPhi=\vU\vV^T/2+\vU'\diag(1/2,0,\ldots,0)\vV'^T$. This is an arbitrary shift.
  \item[(f)] $\lambda \vPhi=-\vU\vV^T/2+\vU'\vV'^T/2$. This is an arbitrary shift.
\end{enumerate}


\begin{figure}[H]
	\centering
	\subfloat[$\lambda \vPhi=\bm{0}$]{\includegraphics[width=2.3in]{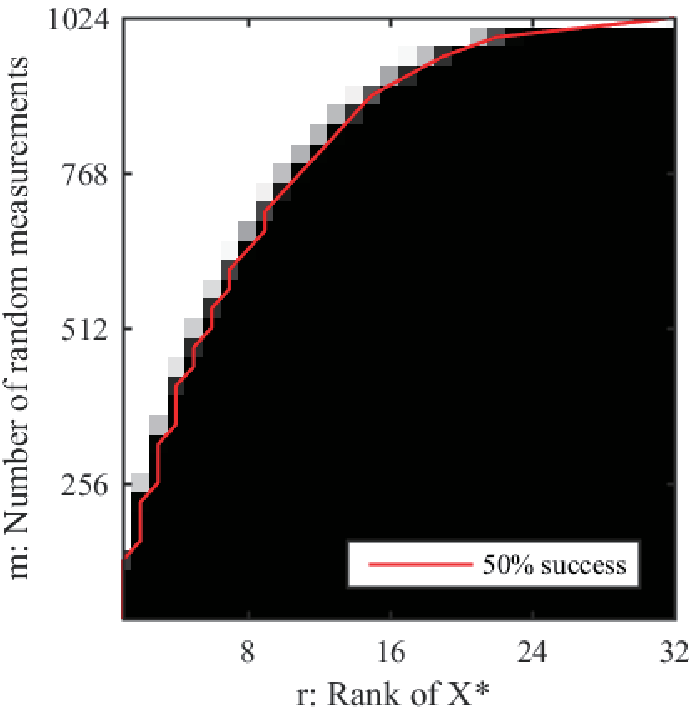}
		\label{fig5_first_case}}
	\hfil
	\subfloat[$\lambda \vPhi= \vU\vV^T/2$]{\includegraphics[width=2.3in]{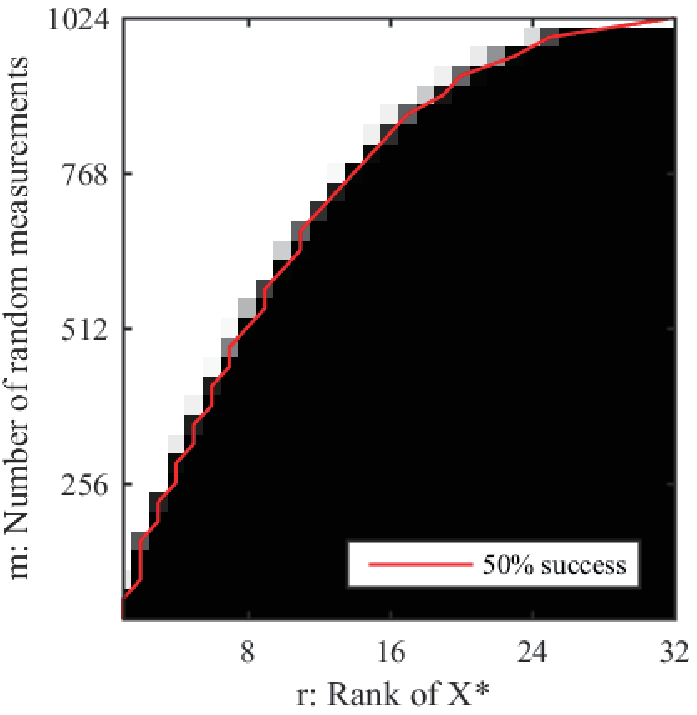}
		\label{fig5_second_case}}
	\hfil
	\subfloat[$\lambda \vPhi=-\vU\vV^T/2$]{\includegraphics[width=2.3in]{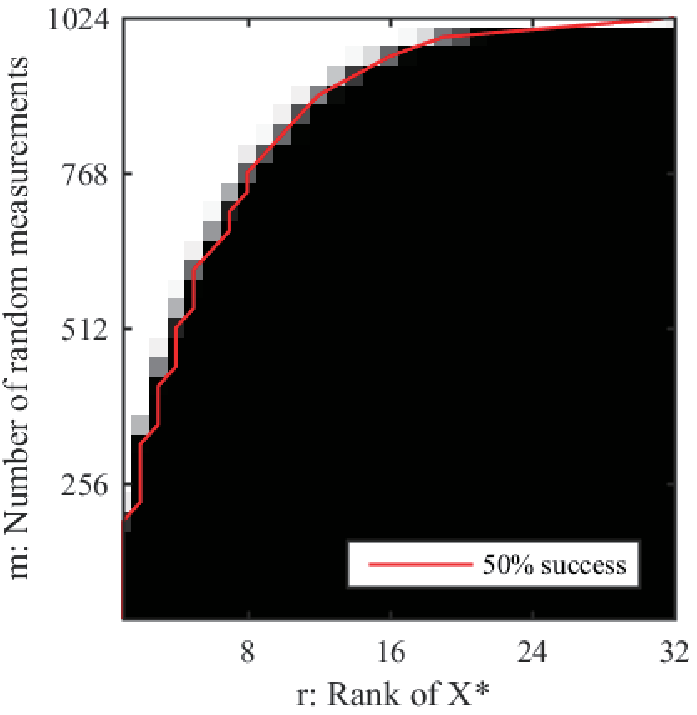}
		\label{fig5_third_case}}
	\hfil
	\subfloat[$\lambda \vPhi=\vU'\vV'^T/2$]{\includegraphics[width=2.3in]{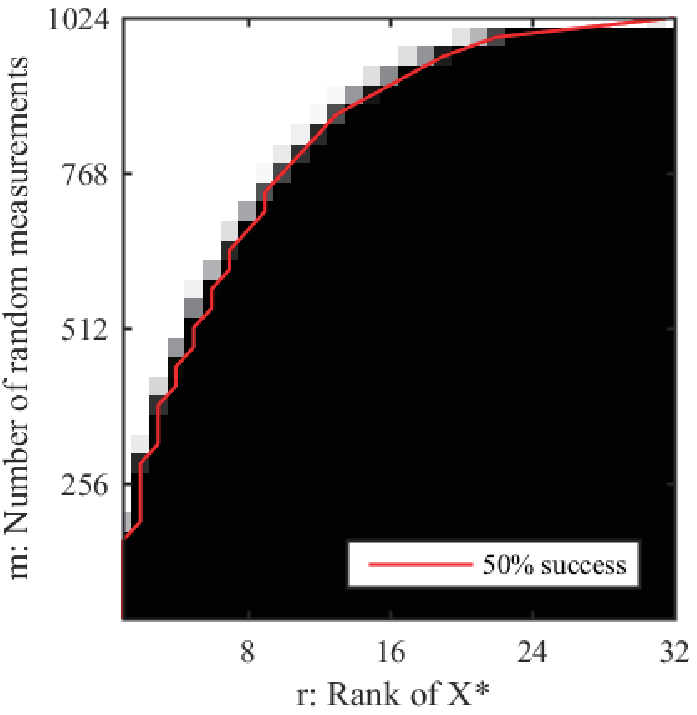}
		\label{fig5_fourth_case}}
	\hfil
	\subfloat[$\lambda \vPhi=\vU\vV^T/2+\vU'\diag(1/2,0,\ldots,0)\vV'^T$]{\includegraphics[width=2.3in]{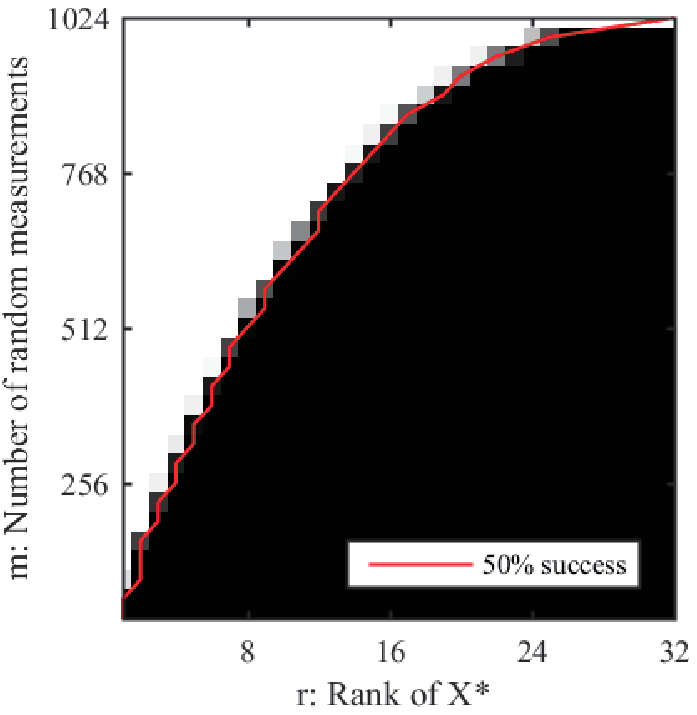}
		\label{fig5_fifth_case}}
	\hfil
	\subfloat[$\lambda \vPhi=-\vU\vV^T/2+\vU'\vV'^T/2$]{\includegraphics[width=2.3in]{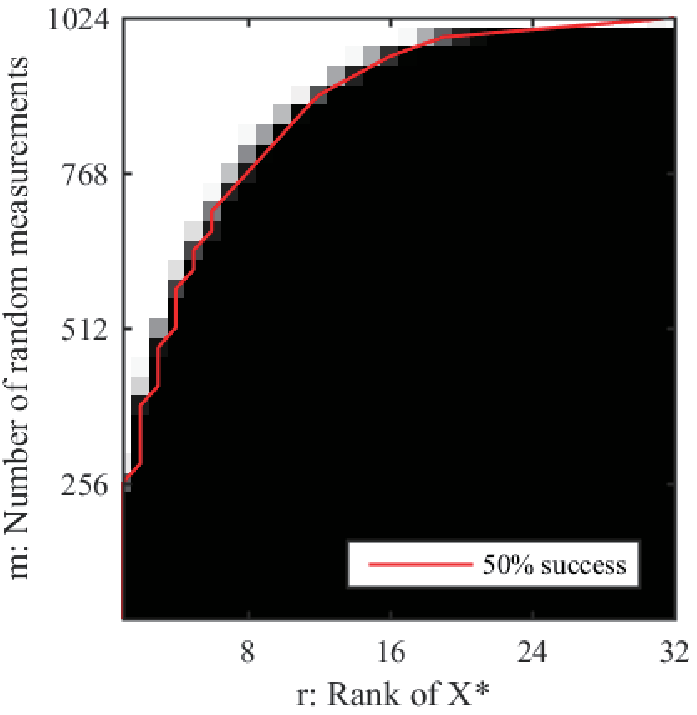}
		\label{fig5_sixth_case}}
	\caption{The phase transition curves of low-rank recovery under different shifts . The brightness of each point reflects the observed probability of success, ranging from black (0\%) to white (100\%).}
	\label{fig: PhaseTransitionofLowRank}
\end{figure}

For reference, we draw Fig. \ref{fig: PhaseTransitionofLowRank}(a) for the classical low-rank recovery procedure without prior information. For shifts on $\mathcal{S}$, Fig. \ref{fig: PhaseTransitionofLowRank}(b) improves the performance a lot while Fig. \ref{fig: PhaseTransitionofLowRank}(c) deteriorates the performance compared to Fig. \ref{fig: PhaseTransitionofLowRank}(a). Comparing Fig. \ref{fig: PhaseTransitionofLowRank}(d) with Fig. \ref{fig: PhaseTransitionofLowRank}(a), we realize that the shift on $\mathcal{S}^{\perp}$ increases the number of measurements. Fig. \ref{fig: PhaseTransitionofLowRank}(e) shows an improved performance for this arbitrary shift. However, Fig. \ref{fig: PhaseTransitionofLowRank}(f) presents an opposite result for another arbitrary shift. All of these experimental results coincide with our theoretical result (Theorem \ref{thm: LowRankCase}) and geometrical interpretation. From these results, we also come to the conclusion that with proper prior information, our approach can have a better performance than the classical low-rank recovery procedure.

\begin{figure}
	\centering
	\includegraphics[width=3.5in]{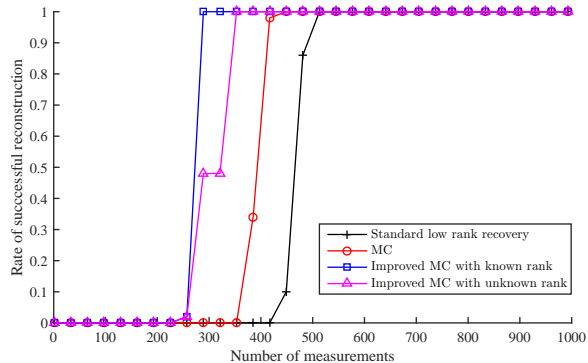}
	\caption{Rate of successful reconstruction v.s. number of measurements for the standard low-rank recovery procedure, MC, improved MC with known rank, and improved MC with unknown rank.}
	\label{fig: PerformanceComparisonLowrank}
\end{figure}

\subsection{Performance Comparisons with Standard Approach for Low-rank Recovery}

In this experiment, we compare the performance of the standard low-rank recovery procedure, MC \eqref{eq: Main_Problem_LowRank}, improved MC with known rank, and improved MC with unknown rank. Here, improved MCs follow the strategies in Section \ref{Guidance} for low-rank matrix recovery.

The original signal $\vX^\star \in \R^{n \times n}$ is generated similarly as in the previous subsection and the measurement matrices $\{\vA^j\}_{j=1}^{m}$ are independent Bernoulli matrices. The observation noise $\vn \in \R^m$ is a random vector whose entries are independently drawn from a zero-mean Gaussian distribution with standard deviation $0.01$. We set the upper bound of the noise vector $\delta=\norm{\vn}_2$.  The prior information is generated as $\vPhi=\vX^\star+\vZ$, where the entries of $\vZ$ are drawn from a zero-mean Gaussian distribution with standard deviation $0.1$. We set $n=32$ and $r=5$. For a fixed $m$, we make $50$ trials and calculate the successful probability as before. We set $\lambda=1/n$ for MC and $\kappa=0.95$ for two improved MCs. The results are shown in Fig. \ref{fig: PerformanceComparisonLowrank}.

Compared to the standard low-rank recovery procedure, all three MCs require less measurements to reconstruct $\vX^\star$ successfully. For the MC approaches, the two improved MCs outperform the normal one and the improved MC with known rank shows a better performance than the one with unknown rank. In summary, our approach can improve the performance of the standard low-rank recovery procedure. The simulation results also illustrate the validity of the guidance for improving prior information.

\section{Conclusion} \label{sec: Conclusion}

This paper has proposed a new approach to recover structured signals with the aid of a similar signal which is known a priori. Theoretical bounds on the number of measurements have been established under sub-Gaussian measurements. Specific structured signals including sparse vectors, block-sparse vectors, and low-rank matrices has also been analyzed. The corresponding geometrical interpretation has been given to illustrate the essence of the proposed approach. Our theoretical results and geometrical interpretation also suggest some strategies to improve the quality of prior information. Numerical simulations are provided to demonstrate the validity of the results.

\appendices
\section{}\label{Appdendix}
\subsection{Proof of Lemma \ref{lm: GaussianWidth_Sparse}}
According to Fact \ref{Prop: BoundGaussWidth}, we have
	\begin{equation*}
	w(\mathcal{T}_f \cap \S^{n-1}) \le \E [\dist(\vg,\mathcal{N}_f)],
	\end{equation*}
	where $\mathcal{N}_f$ denotes the polar of $\mathcal{T}_f$. Since $\bm{0} \notin \partial \norm{\vx^\star}_1 - \lambda \vphi$, it then follows from Fact \ref{Prop: NormalCone} that
	$$ \mathcal{N}_f =\text{cone} \{\partial \norm{\vx^\star}_1 - \lambda \vphi\}. $$
	Thus, by Jensen's inequality, we have
	\begin{equation}\label{optimalboundequation}
	w^2(\mathcal{T}_f \cap \S^{n-1}) \le \E [\dist^2(\vg, \text{cone} \{\partial \norm{\vx^\star}_1 - \lambda \vphi\})].
	\end{equation}
	
	\textbf{Upper bound I.} Fix any $\vg \in \R^n$ and choose any
	$$\vw_0 \in \arg \max \limits_{\vw \in \partial \norm{\vx^\star}_1 - \lambda \vphi} \ip{\vg}{\vw},$$
	then we have, for any $t \geq 0$,
	\isdraft{\begin{equation*}   
		\begin{aligned}
		\dist^2(\vg, \text{\emph{cone}} \{\partial \norm{\vx^\star}_1 - \lambda \vphi\}) & \le \dist^2(\vg, t \cdot (\partial \norm{\vx^\star}_1 - \lambda \vphi)) \\
		& \le \norm{\vg-t\vw_0}_2^2 =\norm{\vg}_2^2 -2t\ip{\vg}{\vw_0}+t^2\norm{\vw_0}_2^2\\
		& = \norm{\vg}_2^2 -2t\max \limits_{\vw \in \partial \norm{\vx^\star}_1 - \lambda \vphi} \ip{\vg}{\vw}+t^2\norm{\vw_0}_2^2 \\
		&\le \norm{\vg}_2^2 -2t\max \limits_{\vw \in \partial \norm{\vx^\star}_1 - \lambda \vphi} \ip{\vg}{\vw}+t^2 \max \limits_{\vw \in \partial \norm{\vx^\star}_1 - \lambda \vphi} \norm{\vw}_2^2.
		\end{aligned}
		\end{equation*}              
	}{
		\begin{equation*}
		\begin{aligned}
		& ~~~ \dist^2(\vg, \text{cone} \{\partial \norm{\vx^\star}_1 - \lambda \vphi\}) \\
		& \le \dist^2(\vg, t \cdot (\partial \norm{\vx^\star}_1 - \lambda \vphi)) \\
		& \le \norm{\vg-t\vw_0}_2^2 =\norm{\vg}_2^2 -2t\ip{\vg}{\vw_0}+t^2\norm{\vw_0}_2^2\\
		& = \norm{\vg}_2^2 -2t\max \limits_{\vw \in \partial \norm{\vx^\star}_1 - \lambda \vphi} \ip{\vg}{\vw}+t^2\norm{\vw_0}_2^2\\
		& \le \norm{\vg}_2^2 -2t\max \limits_{\vw \in \partial \norm{\vx^\star}_1 - \lambda \vphi} \ip{\vg}{\vw}+t^2 \max \limits_{\vw \in \partial \norm{\vx^\star}_1 - \lambda \vphi} \norm{\vw}_2^2.
		\end{aligned}
		\end{equation*}
	}
	Taking expectation on both sides yields
	\isdraft{                   
		\begin{equation} \label{eq: QuadraticBound}
		\begin{aligned}
		\E\[ \dist^2(\vg, t \cdot (\partial \norm{\vx^\star}_1 - \lambda \vphi))\]
		&\le n- 2t \cdot \E \max \limits_{\vw \in \partial \norm{\vx^\star}_1 - \lambda \vphi} \ip{\vg}{\vw} + t^2 \cdot \max \limits_{\vw \in \partial \norm{\vx^\star}_1 - \lambda \vphi} \norm{\vw}_2^2 \\
		& = n- 2t \cdot \sqrt{\frac{2}{\pi}}(n-s) + t^2 \cdot v_1,
		\end{aligned}
		\end{equation}
	}{
		\begin{equation} \label{eq: QuadraticBound}
		\begin{aligned}
		&\E \[ \dist^2(\vg, \text{cone} \{\partial \norm{\vx^\star}_1 - \lambda \vphi\})\] \\
		&\le n- 2t \E \max \limits_{\vw \in \partial \norm{\vx^\star}_1 - \lambda \vphi} \ip{\vg}{\vw} + t^2  \max \limits_{\vw \in \partial \norm{\vx^\star}_1 - \lambda \vphi} \norm{\vw}_2^2 \\
		& = n- 2t \sqrt{\frac{2}{\pi}}(n-s) + t^2 v_1,
		\end{aligned}
		\end{equation}
	}
	where
	$$
	\E \max \limits_{\vw \in \partial \norm{\vx^\star}_1 - \lambda \vphi} \ip{\vg}{\vw}=\E \max \limits_{\vw \in \partial \norm{\vx^\star}_1} \ip{\vg}{\vw} =\sqrt{\frac{2}{\pi}}(n-s).
	$$
	Choosing $t=\sqrt{2/ \pi} (n-s)/v_1$, we achieve the minimum of (\ref{eq: QuadraticBound}) and get the first upper bound.
	
	\textbf{Upper bound II.} Define
	\isdraft{\begin{equation*}
		z(\vg):=   \norm{\vg_{I^c}}_2 \cdot \left(\sign(\vx^\star)+ \frac{\vg_{I^c}}{\norm{\vg_{I^c}}_2} - \lambda \vphi\right) =   \norm{\vg_{I^c}}_2(\sign(\vx^\star)-\lambda \vphi)+\vg_{I^c}.
		\end{equation*}}
	{\begin{align*}
		z(\vg):=  & \norm{\vg_{I^c}}_2 \cdot \left(\sign(\vx^\star)+ \frac{\vg_{I^c}}{\norm{\vg_{I^c}}_2} - \lambda \vphi\right) \\
		=   & \norm{\vg_{I^c}}_2(\sign(\vx^\star)-\lambda \vphi)+\vg_{I^c}.
		\end{align*}}
	Since $ \left\{\sign(\vx^\star)+\vg_{I^c}/\norm{\vg_{I^c}}_2 \right \} \in \partial \norm{\vx^\star}_1$, we have $z(\vg) \in \text{cone} \{\partial \norm{\vx^\star}_1 - \lambda \vphi\}$ and
	\isdraft{       
		\begin{align}
		\E \[ \dist^2(\vg, \text{cone} \{\partial \norm{\vx^\star}_1 - \lambda \vphi\})\]
		&\le \E \norm{\vg-z(\vg)}_2^2 \nonumber\\
		&=\E \norm{\vg_{I}-\norm{\vg_{I^c}}_2\(\sign(\vx^\star) -\lambda \vphi \)}_2^2\nonumber\\
		&=\E \norm{\vg_{I}}_2^2+ \norm{\sign(\vx^\star) -\lambda \vphi}_2^2\E\norm{\vg_{I^c}}_2^2 \label{eq: SparseIndependence}\\
		&=s+(n-s)u_1. \label{eq: DimensionEqualitySparse}
		\end{align}
	}{             
		\begin{align}
		&\quad\E \[ \dist^2(\vg, \text{cone} \{\partial \norm{\vx^\star}_1 - \lambda \vphi\})\]\nonumber\\
		&\le \E \norm{\vg-z(\vg)}_2^2 \nonumber\\
		&=\E \norm{\vg_{I}-\norm{\vg_{I^c}}_2\(\sign(\vx^\star) -\lambda \vphi \)}_2^2\nonumber\\
		&=\E \norm{\vg_{I}}_2^2+ \norm{\sign(\vx^\star) -\lambda \vphi}_2^2\E\norm{\vg_{I^c}}_2^2 \label{eq: SparseIndependence}\\
		&=s+(n-s)u_1.\label{eq: DimensionEqualitySparse}
		\end{align}
	}
	Here \eqref{eq: SparseIndependence} holds because $\vg_{I}$ and $\vg_{I^c}$ are independent. \eqref{eq: DimensionEqualitySparse} follows because $\E \norm{\vg_{I}}_2^2=s$ and $\E\norm{\vg_{I^c}}_2^2=n-s$.
	
	Minimizing the two upper bounds completes the proof.

\subsection{Proof of Lemma \ref{lm: GaussianWidth_Blocksparse}}
	According to Facts \ref{Prop: NormalCone} and \ref{Prop: BoundGaussWidth}, for any $t \ge 0$ and $\bm{0} \notin \partial \norm{\vx^\star}_{2,1} - \lambda \vphi$, we have
	\begin{align*}
	w^2(\mathcal{T}_f \cap \S^{n-1}) &\le \E \[\dist^2(\vg,\mathcal{N}_f)\] \\
	&\le \E \[ \dist^2(\vg, t \cdot (\partial \norm{\vx^\star}_{2,1} - \lambda \vphi)) \],
	\end{align*}
	where $\mathcal{N}_f=\text{cone} \{\partial \norm{\vx^\star}_{2,1} - \lambda \vphi\}$.
	
	\textbf{Upper bound I.} Fix any $\vg \in \R^n$ and choose any
	$$\vw_0 \in \arg \max \limits_{\vw \in \partial \norm{\vx^\star}_{2,1} - \lambda \vphi} \ip{\vg}{\vw},$$
	then we obtain
	\isdraft{               
		\begin{align*}
		\dist^2(\vg, t \cdot (\partial \norm{\vx^\star}_{2,1} - \lambda \vphi))
		& \le \norm{\vg-t\vw_0}_2^2 =\norm{\vg}_2^2 -2t\ip{\vg}{\vw_0}+t^2\norm{\vw_0}_2^2\\
		& = \norm{\vg}_2^2 -2t\max \limits_{\vw \in \partial \norm{\vx^\star}_{2,1} - \lambda \vphi} \ip{\vg}{\vw}+t^2\norm{\vw_0}_2^2\\
		& \le \norm{\vg}_2^2 -2t\max \limits_{\vw \in \partial \norm{\vx^\star}_{2,1} - \lambda \vphi} \ip{\vg}{\vw}+t^2 v_2.
		\end{align*}
	}{                      
		\begin{align*}
		~& ~~\dist^2(\vg, t \cdot (\partial \norm{\vx^\star}_{2,1} - \lambda \vphi)) &\\
		& \le \norm{\vg-t\vw_0}_2^2 =\norm{\vg}_2^2 -2t\ip{\vg}{\vw_0}+t^2\norm{\vw_0}_2^2\\
		& = \norm{\vg}_2^2 -2t\max \limits_{\vw \in \partial \norm{\vx^\star}_{2,1} - \lambda \vphi} \ip{\vg}{\vw}+t^2\norm{\vw_0}_2^2\\
		& \le \norm{\vg}_2^2 -2t\max \limits_{\vw \in \partial \norm{\vx^\star}_{2,1} - \lambda \vphi} \ip{\vg}{\vw}+t^2 v_2.
		\end{align*}}
	Taking expectation on both sides yields
	\isdraft{                   
		\begin{align} \label{eq: QuadraticBound_BlockSparse}
		\E \[ \dist^2(\vg, t \cdot (\partial \norm{\vx^\star}_{2,1} - \lambda \vphi))\]
		&\le n- 2t \E \max \limits_{\vw \in \partial \norm{\vx^\star}_{2,1} - \lambda \vphi} \ip{\vg}{\vw} + t^2  v_2 \nonumber\\
		& = n- 2t w(\partial \norm{\vx^\star}_{2,1}) + t^2 v_2,
		\end{align}
	}{                          
		\begin{align} \label{eq: QuadraticBound_BlockSparse}
		&~~\E \[ \dist^2(\vg, t \cdot (\partial \norm{\vx^\star}_{2,1} - \lambda \vphi))\] \nonumber\\
		&\le n- 2t \E \max \limits_{\vw \in \partial \norm{\vx^\star}_{2,1} - \lambda \vphi} \ip{\vg}{\vw} + t^2  v_2 \nonumber \\
		& = n- 2t w(\partial \norm{\vx^\star}_{2,1}) + t^2 v_2,
		\end{align}
	}
	where $\E\norm{\vg}_2^2=n$ and
	$$\E \max \limits_{\vw \in \partial \norm{\vx^\star}_{2,1}- \lambda \vphi} \ip{\vg}{\vw}=w(\partial \norm{\vx^\star}_{2,1}).$$
	By the subdifferential of the $\ell_1/\ell_2$-norm, we
	\isdraft{
		\begin{align} \label{eq: GaussianWidthBlockSparse}
		w\(\partial \norm{\vx^\star}_{2,1}\)
		&=\E \max \limits_{\vw \in \partial \norm{\vx^\star}_{2,1}} \ip{\vg}{\vw}\nonumber\\
		&= \E \[ \sum \limits_{b \in \mathcal{B}} \ip{\vg_{V_b}}{\frac{\vx^\star_{V_b}}{\norm{\vx^\star_{V_b}}_2}} + \sum \limits_{b \notin \mathcal{B}} \max \limits_{\norm{\bm{\theta}_{V_b}}_2 \le 1} \ip{\vg_{V_b}}{\bm{\theta}_{V_b}}\]\nonumber\\
		&= \sum \limits_{b \notin \mathcal{B}} \E \max \limits_{\norm{\bm{\theta}_{V_b}}_2\le 1} \ip{\vg_{V_b}}{\bm{\theta}_{V_b}} = \sum \limits_{b \notin \mathcal{B}} \E \norm{\vg_{V_b}}_2\nonumber\\
		&= (l-s)\mu_k.
		\end{align}
	}
	{
		\begin{align} \label{eq: GaussianWidthBlockSparse}
		&~~w\(\partial \norm{\vx^\star}_{2,1}\)\nonumber =\E \max \limits_{\vw \in \partial \norm{\vx^\star}_{2,1}} \ip{\vg}{\vw}\nonumber\\
		&= \E \[ \sum \limits_{b \in \mathcal{B}} \ip{\vg_{V_b}}{\frac{\vx^\star_{V_b}}{\norm{\vx^\star_{V_b}}_2}} + \sum \limits_{b \notin \mathcal{B}} \max \limits_{\norm{\bm{\theta}_{V_b}}_2 \le 1} \ip{\vg_{V_b}}{\bm{\theta}_{V_b}}\]\nonumber\\
		&= \sum \limits_{b \notin \mathcal{B}} \E \max \limits_{\norm{\bm{\theta}_{V_b}}_2\le 1} \ip{\vg_{V_b}}{\bm{\theta}_{V_b}}\nonumber = \sum \limits_{b \notin \mathcal{B}} \E \norm{\vg_{V_b}}_2\nonumber\\
		&= (l-s)\mu_k.
		\end{align}
	}
	Combining \eqref{eq: QuadraticBound_BlockSparse} and \eqref{eq: GaussianWidthBlockSparse}, we can achieve the minimum at $t= (l-s)\mu_k/v_2$ and get the first upper bound.
	
	\textbf{Upper bound II.} Define
	\isdraft{                   
		$$
		z(\vg)=\norm{\sum \limits_{b \notin \mathcal{B}}\vg_{V_b}}_2  \( \sum_{b \in \mathcal{B}} \frac{\vx_{V_b}^\star}{\norm{\vx_{V_b}^\star}_2}+\frac{\sum_{b \notin \mathcal{B}}\vg_{V_b}}{\norm{\sum_{b \notin \mathcal{B}}\vg_{V_b}}_2}- \lambda \vphi\)
		=\norm{\sum \limits_{b \notin \mathcal{B}}\vg_{V_b}}_2 \(\sum_{b \in \mathcal{B}} \frac{\vx_{V_b}^\star}{\norm{\vx_{V_b}^\star}_2}-\lambda \vphi\)+\sum_{b \notin \mathcal{B}}\vg_{V_b}.
		$$
	}{                          
		\begin{align*}
		z(\vg) :& =\norm{\sum \limits_{b \notin \mathcal{B}}\vg_{V_b}}_2 \( \sum_{b \in \mathcal{B}} \frac{\vx_{V_b}^\star}{\norm{\vx_{V_b}^\star}_2}+\frac{\sum_{b \notin \mathcal{B}}\vg_{V_b}}{\norm{\sum_{b \notin \mathcal{B}}\vg_{V_b}}_2}- \lambda \vphi \)\\
		& =\norm{\sum \limits_{b \notin \mathcal{B}}\vg_{V_b}}_2 \(\sum_{b \in \mathcal{B}} \frac{\vx_{V_b}^\star}{\norm{\vx_{V_b}^\star}_2}-\lambda \vphi\)+\sum_{b \notin \mathcal{B}}\vg_{V_b}.
		\end{align*}}
	Since $$\sum_{b \in \mathcal{B}} \frac{\vx_{V_b}^\star}{\norm{\vx_{V_b}^\star}_2}+\frac{\sum_{b \notin \mathcal{B}}\vg_{V_b}}{\norm{\sum_{b \notin \mathcal{B}}\vg_{V_b}}_2} \in \partial \norm{\vx^\star}_{2,1},$$
	we have $z(\vg) \in \text{cone} \{\partial \norm{\vx^\star}_{2,1} - \lambda \vphi\}$. Then
	\isdraft{                   
		\begin{align}
		\E \[ \dist^2(\vg,\text{cone} \{\partial \norm{\vx^\star}_{2,1} - \lambda \vphi\})\]
		&\le \E \norm{\vg-z(\vg)}_2^2 \nonumber\\
		&=\E \norm{\sum_{b \in \mathcal{B}}\vg_{V_b}-\norm{\sum \limits_{b \notin \mathcal{B}}\vg_{V_b}}_2 \(\sum_{b \in \mathcal{B}} \frac{\vx_{V_b}^\star}{\norm{\vx_{V_b}^\star}_2}-\lambda \vphi\)}_2^2\nonumber\\
		&=\sum_{b \in \mathcal{B}}\E \norm{\vg_{V_b}}_2^2+ u_2 \cdot \( \sum \limits_{b \notin \mathcal{B}} \E\norm{\vg_{V_b}}_2^2 \)\label{eq: BlockSparseIndependence}\\
		&=k\cdot \(s+(l-s)u_2\).\label{eq: DimensionEqualityBlockSparse}
		\end{align}
	}{                          
		\begin{align}
		&~~\E \[ \dist^2(\vg, \text{cone} \{\partial \norm{\vx^\star}_{2,1} - \lambda \vphi\})\]\nonumber\\
		&\le \E \norm{\vg-z(\vg)}_2^2 \nonumber\\
		&=\E \norm{\sum_{b \in \mathcal{B}}\vg_{V_b}-\norm{\sum \limits_{b \notin \mathcal{B}}\vg_{V_b}}_2 \(\sum_{b \in \mathcal{B}} \frac{\vx_{V_b}^\star}{\norm{\vx_{V_b}^\star}_2}-\lambda \vphi\)}_2^2\nonumber\\
		&=\sum_{b \in \mathcal{B}}\E \norm{\vg_{V_b}}_2^2+ u_2 \cdot \( \sum \limits_{b \notin \mathcal{B}} \E\norm{\vg_{V_b}}_2^2 \)\label{eq: BlockSparseIndependence}\\
		&=k\cdot \(s+(l-s)u_2\).\label{eq: DimensionEqualityBlockSparse}
		\end{align}
	}
	Here \eqref{eq: BlockSparseIndependence} holds because $\sum_{b \in \mathcal{B}}\vg_{V_b}$and $\sum_{b \notin \mathcal{B}}\vg_{V_b}$ are independent. \eqref{eq: DimensionEqualityBlockSparse} follows from $\E \norm{\vg_{V_b}}_2^2=k, \forall~b\in \mathcal{B}$.
	
	Minimizing the two upper bounds completes the proof.

\subsection{Proof of Lemma \ref{lm: GaussianWidth_LowRank}}
	Let $\vG \in \R^{n_1 \times n_2}$ be a Gaussian random matrix with i.i.d. standard Gaussian entries. According to Facts \ref{Prop: NormalCone} and \ref{Prop: BoundGaussWidth}, if $\bm{0} \notin \partial \norm{\vX^\star}_* - \lambda \vPhi$, then for any $t\ge0$,
	\isdraft{
		\begin{equation*} \label{eq: GaussianWidthInequality}
		w^2(\mathcal{T}_f \cap \S^{n_1 n_2 -1}) \le \E[\dist^2 (\vG,\text{cone}\{\partial \norm{\vX^\star}_* -\lambda \vPhi\})]
		\le \E[\dist^2 (\vG,t \cdot (\partial \norm{\vX^\star}_* -\lambda \vPhi))],
		\end{equation*}
	}
	{
		\begin{align*}
		w^2(\mathcal{T}_f \cap \S^{n_1 n_2 -1}) &\le \E[\dist^2 (\vG,\text{cone}\{\partial \norm{\vX^\star}_* -\lambda \vPhi\})] \\
		&\le \E[\dist^2 (\vG,t \cdot (\partial \norm{\vX^\star}_* -\lambda \vPhi))],
		\end{align*}
	}
	where $\dist^2(\vX,\vY)=\norm{\vX-\vY}_F^2$.
	
	\textbf{Upper bound I.} Fix any $\vG \in \R^{n_1 \times n_2}$ and choose any
	$$
	\vZ_0 \in \arg \max \limits_{\vZ \in \partial \norm{\vX^\star}_* -\lambda \vPhi} \ip{\vG}{\vZ},
	$$
	then we have
	\isdraft
	{           
		\begin{align*}
		\dist^2 (\vG,t \cdot (\partial \norm{\vX^\star}_* -\lambda \vPhi)) &\le \norm{\vG-t\vZ_0}_F^2
		=\norm{\vG}_F^2 -2t \ip{\vG}{\vZ_0} + t^2 \norm{\vZ_0}_F^2 \\
		&\le \norm{\vG}_F^2 -2t \max \limits_{\vZ \in \partial \norm{\vX^\star}_* -\lambda \vPhi} \ip{\vG}{\vZ} + t^2 \norm{\vZ_0}_F^2.
		\end{align*}
	}{                  
		\begin{align*}
		~&~~\dist^2 (\vG,t \cdot (\partial \norm{\vX^\star}_* -\lambda \vPhi)) \le \norm{\vG-t\vZ_0}_F^2 \\
		&=\norm{\vG}_F^2 -2t \ip{\vG}{\vZ_0} + t^2 \norm{\vZ_0}_F^2 \\
		&\le \norm{\vG}_F^2 -2t \max \limits_{\vZ \in \partial \norm{\vX^\star}_* -\lambda \vPhi} \ip{\vG}{\vZ} + t^2 \norm{\vZ_0}_F^2.
		\end{align*}}
	Taking expectations on both sides, we obtain
	\isdraft{          
		\begin{align*}
		w^2(\mathcal{T}_f \cap \S^{n_1 n_2 -1})
		& \le \E \norm{\vG}_F^2 -2t \cdot w(\partial \norm{\vX^\star}_*) +t^2 \max \limits_{\vZ \in \partial \norm{\vX^\star}_* -\lambda \vPhi} \norm{\vZ}_F^2 \\
		& \le n_1n_2 -2t \cdot w(\partial \norm{\vX^\star}_*) +t^2 v_3,
		\end{align*}
	}{                 
		\begin{align*}
		&~~w^2(\mathcal{T}_f \cap \S^{n_1 n_2 -1})\\
		& \le \E \norm{\vG}_F^2 -2t \cdot w(\partial \norm{\vX^\star}_*) +t^2 \max \limits_{\vZ \in \partial \norm{\vX^\star}_* -\lambda \vPhi} \norm{\vZ}_F^2 \\
		& \le n_1n_2 -2t \cdot w(\partial \norm{\vX^\star}_*) +t^2 v_3,
		\end{align*}}
	where $\E \norm{\vG}_F^2=n_1n_2$ and
	$$
	w(\partial \norm{\vX^\star}_*) = \E \max \limits_{\vZ \in \partial \norm{\vX^\star}_*} \ip{\vG}{\vZ} = \E \max \limits_{\vZ \in \partial \norm{\vX^\star}_* -\lambda \vPhi} \ip{\vG}{\vZ}.
	$$
	According to \cite[Lemma 1]{foygel2014corrupted}, we have
	$$
	w(\partial \norm{\vX^\star}_*) \ge \frac{4}{27}(n_2-r)\sqrt{n_1-r}.
	$$
	Choosing $t=w(\partial \norm{\vX^\star}_*)/v_3$, we achieve the minimum
	\isdraft{               
		\begin{equation*}
		w^2(\mathcal{T}_f \cap \S^{n_1 n_2 -1}) \le n_1 n_2\( 1- \frac{n_2}{v_3} \(\frac{4}{27}\)^2 \(1-\frac{r}{n_1}\)\(1-\frac{r}{n_2}\)^2 \).
		\end{equation*}
	}{                      
		\begin{multline*}
		w^2(\mathcal{T}_f \cap \S^{n_1 n_2 -1}) \\
		\le n_1 n_2\( 1- \frac{n_2}{v_3} \(\frac{4}{27}\)^2 \(1-\frac{r}{n_1}\)\(1-\frac{r}{n_2}\)^2 \).
		\end{multline*}}
	
	\textbf{Upper bound II.} Define
	\isdraft{               
		$$
		Z(\vG)=\norm{\mathcal{P}_{\mathcal{S}^\bot}(\vG)}\(\vU\vV^T+\frac{\mathcal{P}_{\mathcal{S}^\bot}(\vG)}{\norm{\mathcal{P}_{\mathcal{S}^\bot}(\vG)}}-\lambda \vPhi\)=\norm{\mathcal{P}_{\mathcal{S}^\bot}(\vG)}(\vU\vV^T-\lambda \vPhi) +\mathcal{P}_{\mathcal{S}^\bot}(\vG).
		$$
	}{                      
		\begin{align*}
		Z(\vG) & =\norm{\mathcal{P}_{\mathcal{S}^\bot}(\vG)}\(\vU\vV^T+\frac{\mathcal{P}_{\mathcal{S}^\bot}(\vG)}{\norm{\mathcal{P}_{\mathcal{S}^\bot}(\vG)}}-\lambda \vPhi\) \\
		& =\norm{\mathcal{P}_{\mathcal{S}^\bot}(\vG)}(\vU\vV^T-\lambda \vPhi) +\mathcal{P}_{\mathcal{S}^\bot}(\vG).
		\end{align*}
	}
	Since $\vU\vV^T+{\mathcal{P}_{\mathcal{S}^\bot}(\vG)}/{\norm{\mathcal{P}_{\mathcal{S}^\bot}(\vG)}} \in \partial \norm{\vX^\star}_*$, we have $Z(\vG) \in \text{cone}\{\partial \norm{\vX^\star}_*-\lambda\vPhi\}$. Then
	\isdraft{               
		\begin{align}
		w^2(\mathcal{T}_f \cap \S^{n_1 n_2 -1})
		&\le \E \norm{\vG-Z(\vG)}_F^2 \nonumber \\
		&= \E \norm{\mathcal{P}_{\mathcal{S}}(\vG)-\norm{\mathcal{P}_{\mathcal{S}^\bot}(\vG)}(\vU\vV^T-\lambda \vPhi)}_F^2 \nonumber\\
		&= \E \norm{\mathcal{P}_{\mathcal{S}}(\vG)}_F^2 + \norm{\vU\vV^T-\lambda \vPhi}_F^2 \E \norm{\mathcal{P}_{\mathcal{S}^\bot}(\vG)}^2 \label{eq: ProjectionIndependence}\\
		&= r(n_1+n_2-r) + u_3 \E \norm{\mathcal{P}_{\mathcal{S}^\bot}(\vG)}^2. \label{eq: DimensionEquality}
		\end{align}
	}{                      
		\begin{align}
		&~~w^2(\mathcal{T}_f \cap \S^{n_1 n_2 -1}) \nonumber\\
		&\le \E \norm{\vG-Z(\vG)}_F^2 \nonumber \\
		&= \E \norm{\mathcal{P}_{\mathcal{S}}(\vG)-\norm{\mathcal{P}_{\mathcal{S}^\bot}(\vG)}(\vU\vV^T-\lambda \vPhi)}_F^2 \nonumber\\
		&= \E \norm{\mathcal{P}_{\mathcal{S}}(\vG)}_F^2 + \norm{\vU\vV^T-\lambda \vPhi}_F^2 \E \norm{\mathcal{P}_{\mathcal{S}^\bot}(\vG)}^2 \label{eq: ProjectionIndependence}\\
		&= r(n_1+n_2-r) + u_3 \E \norm{\mathcal{P}_{\mathcal{S}^\bot}(\vG)}^2. \label{eq: DimensionEquality}
		\end{align}}
	Here \eqref{eq: ProjectionIndependence} holds because $\mathcal{P}_{\mathcal{S}}(\vG)$ and $\mathcal{P}_{\mathcal{S}^\bot}(\vG)$ are independent. \eqref{eq: DimensionEquality} follows because  $ \E \norm{\mathcal{P}_{\mathcal{S}}(\vG)}_F^2=r(n_1+n_2-r)$.
	According to \cite[equation (87)]{chandrasekaran2012convex}, we have
	\begin{equation*}
	\E \norm{\mathcal{P}_{\mathcal{S}^\bot}(\vG)}^2 \le (\sqrt{n_1-r}+\sqrt{n_2-r})^2+2.
	\end{equation*}
	Therefore,
	\isdraft{               
		\begin{equation*}
		w^2(\mathcal{T}_f \cap \S^{n_1 n_2 -1}) \le r(n_1+n_2-r) + u_3 \((\sqrt{n_1-r}+\sqrt{n_2-r})^2+2\).
		\end{equation*}
	}{                      
		\begin{multline*}
		w^2(\mathcal{T}_f \cap \S^{n_1 n_2 -1}) \le r(n_1+n_2-r) +\\ u_3 \((\sqrt{n_1-r}+\sqrt{n_2-r})^2+2\).
		\end{multline*}}
	
	Combining the two bounds completes the proof.

%
\ifCLASSOPTIONcaptionsoff
  \newpage
\fi



\bibliographystyle{IEEEtran}
\bibliography{IEEEabrv,ref}
\end{document}